\documentclass[times,12pt]{article}
\usepackage[margin=1in]{geometry}
\usepackage[hidelinks]{hyperref}
\usepackage{lineno}
\usepackage{amsmath, amssymb}
\usepackage{hyperref}
\usepackage{graphicx}
\usepackage[round, comma]{natbib}
\usepackage{psfrag}
\usepackage[lofdepth,lotdepth]{subfig}
\usepackage{mathtools}
\usepackage{multirow, authblk}
\usepackage{multicol}
\usepackage{bm}
\usepackage{threeparttable, tablefootnote}
\usepackage{xcolor}
\usepackage{soul}
\extrafloats{100}
\usepackage{subfig}
\usepackage{chngcntr}
\usepackage{tabularx}
\newcommand{\GG}[1]{}
\hypersetup{
    colorlinks,
    linkcolor={red!80!black},
    citecolor={blue!50!black},
    urlcolor={blue!80!black}
}

\begin{document}

\begin{center}
    \textbf{Callisto's atmosphere: First evidence for H$_2$ and constraints on H$_2$O} \\
\end{center}

\noindent S.R. Carberry Mogan$^{1,2,3,4,5,*}$, O.J. Tucker$^{6}$, R.E. Johnson$^{2,7}$, L. Roth$^{5}$, J. Alday$^{8, 9}$, A. Vorburger$^{4}$, P.~Wurz$^{4}$,  A. Galli$^{4}$, H.T. Smith$^{10}$, B. Marchand$^{3}$, A.V. Oza$^{4, 11}$\\

\noindent $^1$UC Berkeley, California, USA; $^2$NYU, New York, USA; $^3$NYU Abu Dhabi, Abu Dhabi, UAE; $^4$University of Bern, Bern, Switzerland; $^5$KTH Royal Institute of Technology, Stockholm, Sweden; $^6$NASA Goddard Space Flight Center, Greenbelt, USA; $^7$University of Virginia, Charlottesville, USA; $^8$University of Oxford, Oxford, England; $^9$Open University, Milton Keynes, England; $^{10}$Johns Hopkins University APL, Laurel, USA; $^{11}$NASA JPL, Pasadena, USA \\

\noindent $^*$Corresponding author: Shane R. Carberry Mogan (CarberryMogan@berkeley.edu)

\section*{Abstract}

We explore the parameter space for the contribution to Callisto's H corona observed by the Hubble Space Telescope \citep{roth2017a} from sublimated H$_2$O and radiolytically produced H$_2$ using the Direct Simulation Monte Carlo (DSMC) method. The spatial morphology of this corona produced via photo- and magnetospheric electron impact-induced dissociation is described by tracking the motion of and simulating collisions between the hot H atoms and thermal molecules including a near-surface O$_2$ component. Our results indicate that sublimated H$_2$O produced from the surface ice, whether assumed to be intimately mixed with or distinctly segregated from the dark non-ice or ice-poor regolith, cannot explain the observed structure of the H corona. On the other hand, a global H$_2$ component can reproduce the observation, and is also capable of producing the enhanced electron densities observed at high altitudes by \textit{Galileo}'s plasma-wave instrument \citep{gurnett1997, gurnett2000}, providing the first evidence of H$_2$ in Callisto's atmosphere. The range of H$_2$ surface densities explored, under a variety of conditions, that are consistent with these observations is $\sim$(0.4--1)$\times$10$^8$~cm$^{-3}$. The simulated H$_2$ escape rates and estimated lifetimes suggest that Callisto has a neutral H$_2$ torus. We also place a rough upper limit on the peak H$_2$O number density ($\lesssim$10$^8$~cm$^{-3}$), column density ($\lesssim$10$^{15}$~cm$^{-2}$), and sublimation flux ($\lesssim$10$^{12}$ cm$^{-2}$ s$^{-1}$), all of which are 1--2 orders of magnitude less than that assumed in previous models. Finally, we discuss the implications of these results, as well as how they compare to Europa and Ganymede.

\section*{Plain Language Summary}

The surface and atmosphere of Callisto, the outermost Galilean moon of Jupiter, are not well understood. Although water ice is a significant fraction of its bulk composition, there is no consensus on the amount of surface ice nor how that correlates with the amount of atmospheric water vapor produced via sublimation. Similarly, although irradiation of the icy surface by the plasma trapped in Jupiter's magnetic field is expected to release O$_2$ and H$_2$ as well as directly eject H$_2$O into the atmosphere, only near-surface O$_2$ and trace extended H components have been observed by the Hubble Space Telescope, while H$_2$O and H$_2$ have not. By simulating the motion of these four species in Callisto's atmosphere, we estimated the contributions to the extended H atmosphere via dissociation of H$_2$O and H$_2$. Using sublimation rates suggested in the literature, H$_2$O produces too much H near the subsolar point and too little closer to the terminator to reproduce the observation. On the other hand, a more global tenuous H$_2$ component can explain the Hubble observation, as well as earlier observations made by the Galileo spacecraft of a highly extended ionosphere. This provides the first evidence for H$_2$ in Callisto's atmosphere.

\section{Introduction} \label{intro}

As the most geologically primitive of the icy Galilean satellites, Callisto has the least well understood atmosphere, limiting our understanding of the evolution of the objects in this important system, soon to be the subject of multiple new spacecraft observations. There is no consensus on the state of water ice on its surface nor how that correlates with the production of atmospheric water vapor. Similarly, although radiolysis is likely the primary source of O$_2$ in Callisto's atmosphere \citep{cunningham2015}, the concomitant H$_2$ component has not yet been identified. Forthcoming exploration of Callisto by ESA's JUpiter ICy moons Explorer (JUICE) (e.g., \citealt{galli2022}), NASA's Europa Clipper, and CNSA's planned Gan De can help resolve such issues. With these missions as motivation we expand on our earlier simulations of Callisto's atmosphere \citep{carberrymogan2021} using the observation of its H corona \citep{roth2017a} to examine the limits of both its H$_2$O and H$_2$ components.

The importance of the interrelated surface and atmospheric processes at Callisto are exemplified by its O$_2$ atmosphere \citep{cunningham2015} and its H corona \citep{roth2017a}. The former is likely produced via radiolysis in Callisto's icy surface (e.g., \citealt{johnson1990}) as well as, to a much less extent, via a series of photochemical reactions of sublimated H$_2$O (e.g., \citealt{yung1977}). Since O$_2$ does not freeze out on the surface, even on the night side, it permeates the porous regolith and enriches the atmosphere limited primarily by reactions in the regolith and gas-phase ionizing and dissociative processes. An H corona was detected by the Hubble Space Telescope (HST) \citep{roth2017a}, which was suggested to be produced via photolytic or electron impact dissociation of sublimated H$_2$O or, possibly, radiolytically produced H$_2$, with a very small contribution due to direct sputtering from its icy surface. \cite{carberrymogan2021}, hereafter referred to as ``CM21,'' showed that even though radiolytically produced H$_2$ can have a peak density orders of magnitude less than that of the sublimated H$_2$O component, it can be the primary producer of H near and beyond the terminator. We follow up on that study by using the morphology of the observed H corona to explore the parameter space of its source to place constraints on the very uncertain amounts of sublimated H$_2$O and radiolytically produced H$_2$. Before describing the modeling (Section~\ref{model}), the results (Section~\ref{results}), and the implications (Section~\ref{discussion}), we first review below the observations of Callisto's atmosphere and ionosphere, of water ice on the surface and its relation to the production of water vapor, as well as the sources and corresponding structure of our proposed H$_2$ atmospheric component.

The tenuous CO$_2$ atmosphere observed by \textit{Galileo} \citep{carlson1999} was suggested to be global, and radiolysis was suggested to be one of its possible sources. In addition, \textit{Galileo} radio occultations indicated the presence of a substantial ionosphere located at Callisto's terminator \citep{kliore2002}. Analogous to the O$_2$ atmosphere on Europa inferred from oxygen emissions \citep{hall1995}, the ionosphere was suggested to be sourced by a collisional O$_2$ atmosphere about 2 orders of magnitude more dense than that of the observed CO$_2$, $\sim$4$\times$10$^8$ cm$^{-3}$. This substantial, near-surface ionosphere was only seen at western elongation: when the trailing hemisphere (TH) of Callisto was simultaneously illuminated by the Sun and bombarded by the co-rotating Jovian magnetospheric plasma; i.e., when Callisto's day-side and co-rotating plasma ``ram-side'' hemispheres were aligned. On the other hand, a highly extended ionospheric plasma was detected at eastern elongation \citep{gurnett1997, gurnett2000}: when Callisto's leading hemisphere (LH), which is opposite its ram-side hemisphere, was illuminated. That is, during the C3 and C10 flybys, with a closest approach (C/A) of 1129 km (1.47 $R_C$, where $R_C = 2410$ km is the radius of Callisto; \citealt{gurnett1997}) and 535 km (1.22 $R_C$; \citealt{gurnett2000}), respectively, as \textit{Galileo} passed through the wake downstream of the co-rotating Jovian magnetospheric plasma, electron densities orders of magnitude larger than those expected from the background plasma, $<$1 cm$^{-3}$ \citep{gurnett2000}, were inferred from its plasma-wave measurements. These electron densities were comparable to those seen near Ganymede \citep{barth1997}, the source of which was suggested to be an extended neutral component, suggesting a similar feature is also present at Callisto. During the C22 flyby, in which \textit{Galileo} passed by the night-side through the plasma wake at western elongation with a C/A of 2299 km (1.95 $R_C$) an enhanced extended plasma was not observed \citep{gurnett2000}, although this was the orbit during which one of the largest electron densities, $\sim$(0.8--1.5)$\times$10$^4$ cm$^{-3}$, were detected by \cite{kliore2002} near Callisto's surface, $\sim$8--28 km.

Observations by the HST-Space Telescope Imaging Spectrograph (STIS) were initially unable to detect ultraviolet (UV) auroral emissions caused by magnetospheric electron impact ionization of Callisto's atmosphere \citep{strobel2002}. The authors suggested a possible explanation for this: the Jovian magnetospheric plasma is largely diverted by Callisto's ionosphere. As a result, penetration of magnetospheric electrons into Callisto's atmosphere could not be the source of its near-surface ionosphere. Subsequent atomic oxygen emissions were detected using the HST-Cosmic Origins Spectrograph, which were suggested to be induced by photoelectron impacts in a near-surface, O$_2$-dominated atmosphere when Callisto's LH was illuminated \citep{cunningham2015}. The derived O$_2$ column density, $\sim$4$\times$10$^{15}$ cm$^{-2}$, is an order of magnitude less than that inferred by \cite{kliore2002} when Callisto's TH was illuminated, $\sim$4$\times$10$^{16}$~cm$^{-2}$. Finally, the HST/STIS observations made by \cite{strobel2002} were recently revisited and faint emissions above Callisto's limb were detected \citep{roth2017a}, likely originating from resonant scattering by an H corona.

Even after several decades of research and multiple spacecraft missions, the state of surface water ice at Callisto and the concomitant production of atmospheric water vapor are not well constrained. \textit{Galileo} measurements of Callisto's gravity field suggest 40$\%$ of its bulk composition is ice \citep{anderson1997}, with later analyses yielding mass fractions of 49--55$\%$ ice \citep{spohn2003, kuskov2005}. However, estimates for surficial coverage of ice on its LH and TH range from only 5--30$\%$ \citep{pilcher1972, mandeville1980, clarkmccord1980, spencer1987a, roush1990, mccord1998} with an additional 0--10$\%$ bound water \citep{clarkmccord1980}, while the remainder is a relatively dark silicate and/or carbonaceous material \citep{mccord1998}. Whereas these ranges only refer to the visible ice patches on its surface, others have suggested that ice intimately mixed with the non-ice surface material (i.e., ``dirty ice'') is more abundant with weight fractions ranging from 20--90 wt$\%$ \citep{clark1980, roush1990, calvin1991, mccord1998} with an additional 0--10 wt$\%$ bound water \citep{clark1980}, although much lower weight fractions have also been suggested (e.g., $\sim$4--6 wt$\%$ \citealt{spencer1987a, spencer1987b}). See CM21 for a more in-depth review of the water ice-related observations and models at Callisto. 

Thus it seems that dark material containing adsorbed H$_2$O and relatively sparse ice patches are both present to some extent on Callisto's surface. However, until this study, relevant to modeling Callisto's atmosphere only the former has been considered in which the ice and non-ice or ice-poor material are intimately mixed and H$_2$O sublimates at Callisto's warm day-side temperatures producing a locally relatively dense atmospheric component (e.g., \citealt{liang2005, vorburger2015, hartkorn2017}, CM21). On the other hand, when these materials are segregated from each other (e.g., \citealt{spencer1987b}), the areal coverage and local temperatures of the ice patches will primarily determine the net H$_2$O sublimation rate.

Although gas-phase H$_2$O has not been detected, the recent Hubble observation of an H corona at Callisto \citep{roth2017a} presents a way forward. Analogous to models of Ganymede's atmosphere \citep{marconi2007}, \cite{roth2017a} suggested dissociation of H$_2$O is the main source for the H on the day-side with a possible contribution from H$_2$ near the terminator. The H corona was initially thought to be larger at eastern elongation than at western elongation due to the temperature differences of the LH and TH \citep{roth2017a}. However, the weaker signal from the H corona at western elongation was later found to be affected by absorption in the geocorona \citep{alday2017}. Because of this issue, herein we solely focus on the H observed at eastern elongation, which was negligibly affected.

Using forward models with and without an H corona, the eastern elongation data were in good agreement with a globally symmetric neutral H component (\citealt{roth2017a}, Fig.~4 therein), which produces a peak line-of-sight (LOS) column density at the terminator. CM21 showed that even if the H$_2$O sublimates at the warm day-side surface temperatures, its production of H near the terminator region is negligible and completely dominated by photodissociation of H$_2$. Indeed this is the case near and beyond Callisto's limb even when the peak density of H$_2$ is orders of magnitude less dense than that of H$_2$O (CM21, Fig.~5 therein) because of its global extent and relatively large scale height. This spatial distribution of the H corona at eastern elongation will be used below to help constrain both the H$_2$O and H$_2$ content of Callisto's atmosphere. 

In addition to the above uncertainties for water ice on the surface and its relation to the production of water vapor, estimating the radiolytic source rate for H$_2$ at Callisto is extremely difficult due to the lack of observational constraints and the uncertainty of the surface composition as well as its exposure to the local plasma environment (e.g., \citealt{galli2022} and references therein). This uncertainty is exacerbated as H$_2$ can be radiolytically produced from hydrated sulfur \citep{cartwright2020} and hydrocarbons \citep{mccord1997, mccord1998} in Callisto's non-ice or ice-poor material in addition to being produced from the ice, including from any carbonic acid therein (e.g., \citealt{johnson2004}). Moreover, the very energetic particles can penetrate the non-ice or ice-poor regolith overlying the more ice-rich surface, thereby producing H$_2$ in the ice, which in turn can diffuse through this lag deposit that insulates the underlying ice inhibiting sublimation. Finally, H$_2$ is also a \textit{direct} dissociation product of water ice (e.g., \citealt{teolis2017}) and vapor (e.g., \citealt{itikawa2005, huebner2015}) and can be formed following proton implantation in the surface (e.g., \citealt{tucker2019, tucker2021}).

H$_2$ has a large scale height ($\sim$250--550 km at Callisto's surface temperatures, $\sim$80--167 K); and its escape fraction is small (e.g., \citealt{carberrymogan2020}, hereafter referred to as ``CM20,'' CM21), as are the photo- and electron-impact destruction rates (Table~\ref{tab:reactions} in Appendix~\ref{app:reactions}), so that most H$_2$ produced returns to the surface. Since these returning molecules do not adsorb efficiently (e.g., \citealt{sandford1993, acharyya2014}), they will permeate the porous regolith, in which reaction rates are also negligible allowing it to thermally desorb back into the atmosphere, where it accumulates. In this way, even a relatively small H$_2$ source rate can produce a steady-state, collisional atmospheric component (e.g., CM20, CM21).

In 1D models, CM20 assumed the H$_2$ surface density was related stochiometrically to the observed O$_2$, whereas \cite{liang2005} assumed H$_2$ was solely produced by dissociative recombination of H$_2$O$^+$ and is thus strongly correlated with the sublimated water vapor (Fig. 2 therein). However, to explain the spatial profile of the H corona, a 2D model is required. Such a model of Callisto's atmosphere containing H$_2$O and O$_2$, as well as a range of thermally desorbed H$_2$ assumed to be produced by the mechanisms discussed above was implemented in CM21. Below, that work is extended to account for the fate of the H produced from H$_2$O and H$_2$. Given the limited information about Callisto's atmosphere and surface, accounting for the spatial distribution of the H corona is shown to help constrain the surface source rates of H$_2$O and H$_2$, as well as demonstrate the influence of collisions.

\section{Numerical Method} \label{model}

\subsection{The Direct Simulation Monte Carlo (DSMC) Method}

The direct simulation Monte Carlo (DSMC) method \citep{bird1994} simulates macroscopic gas dynamics via molecular kinetics. By implicitly solving the Boltzmann equation, it can be used to model dense fluids, rarefied gasses, and the \textit{transition} from the former to the latter. Thus it is ideally suited to simulate Callisto's atmosphere as it transitions from thermal equilibrium near the surface to the non-equilibrium rarefied regime (e.g., the exosphere). We have successfully applied the DSMC method to Callisto's atmosphere in earlier studies (CM20, CM21, \citealt{carberrymogan2021b}). Here it is used to simulate Callisto's atmosphere composed of sublimated H$_2$O, radiolytically produced H$_2$ and O$_2$ components, and H produced from H$_2$O and H$_2$ via interactions with photons and magnetospheric electrons in a 2D axisymmetric spherical domain. 

The DSMC method simulates stochastic microscopic processes via computational particles, each of which represents a large number of real atoms or molecules. As these particles traverse physical space they are influenced by gravitational forces, binary collisions, and interactions with photons and magnetospheric electrons, and their motion is tracked using a 4th-order Runge Kutta integration. The 2D spherical grid in which these particles move is decomposed into cells that vary along radial and subsolar latitudinal (SSL) axes (Fig. \ref{fig:DSMC_grid} in Appendix \ref{app:grid}).

Elastic collisions between thermal particles are calculated using the variable hard sphere (VHS) model \citep{bird1994} using the parameters listed in CM21 (Appendix D therein). Collisions between hot H atoms and thermal species are calculated using the model in \cite{lewkow2014}. When calculating collisions between particles representing a different number of atoms or molecules we implement the technique described by \cite{miller1994}.

A DSMC simulation is temporally discretized into time-steps, $\Delta t$. Nascent H particles will initially move at velocities relative to the excess energy of the reaction that produces them (Table~\ref{tab:reactions} in Appendix~\ref{app:reactions}), and these are much faster than the typical speeds of the thermal species. Therefore, we implement sub-time-steps, $\Delta t_\mathrm{sub}$, for the H particles. That is, H particles will move $\Delta t / \Delta t_\mathrm{sub}$ times before the thermal species then move over $\Delta t$. Moreover, collision-based calculations between H and the thermal species will occur over each $\Delta t_\mathrm{sub}$, while such calculations between thermal species only occur over $\Delta t$.

The DSMC simulations presented herein are run for a long enough duration, on the order of Callisto's orbital period, $t_\mathrm{orb}$ = 1.44$\times$10$^6$ s, such that the distribution of particles and their corresponding characteristics yield steady-state macroscopic properties, such as density, temperature, and escape rates. Steady-state is determined through periodic sampling and averaging of the flow field. Upon reaching steady-state, the simulations are run for several more $t_\mathrm{orb}$ with more samples taken to reduce the statistical noise inherent in the resultss, see Fig. \ref{fig:LOS_Column_StdDev} in Appendix \ref{app:stat_noise}.

\subsection{Surface Temperature} \label{sec:surf_temp}

\begin{figure}[ht!]
    \centering
    \subfloat[]{\includegraphics[width=0.67\textwidth]{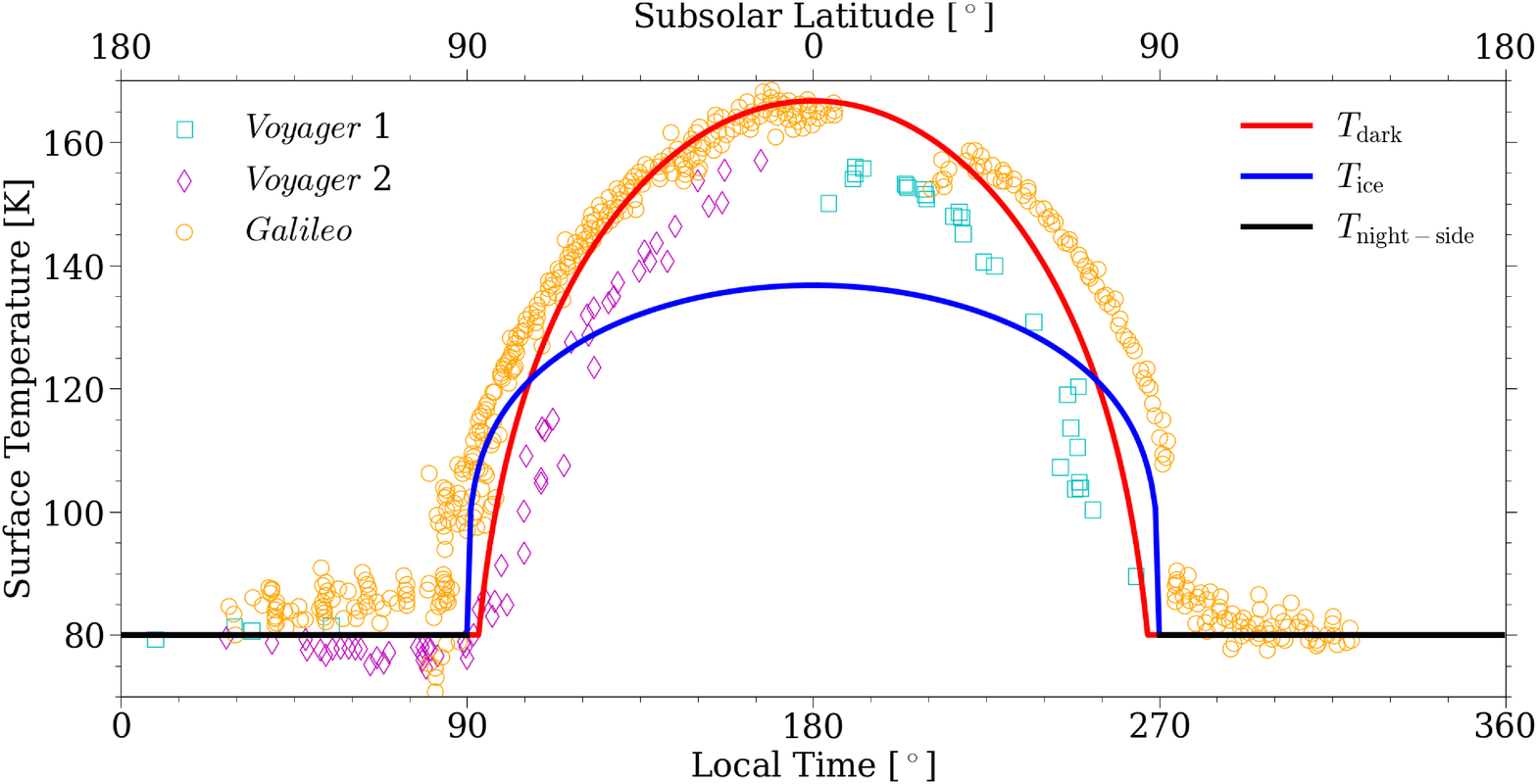}}
    \hfill
    \subfloat[]{\includegraphics[width=0.33\textwidth]{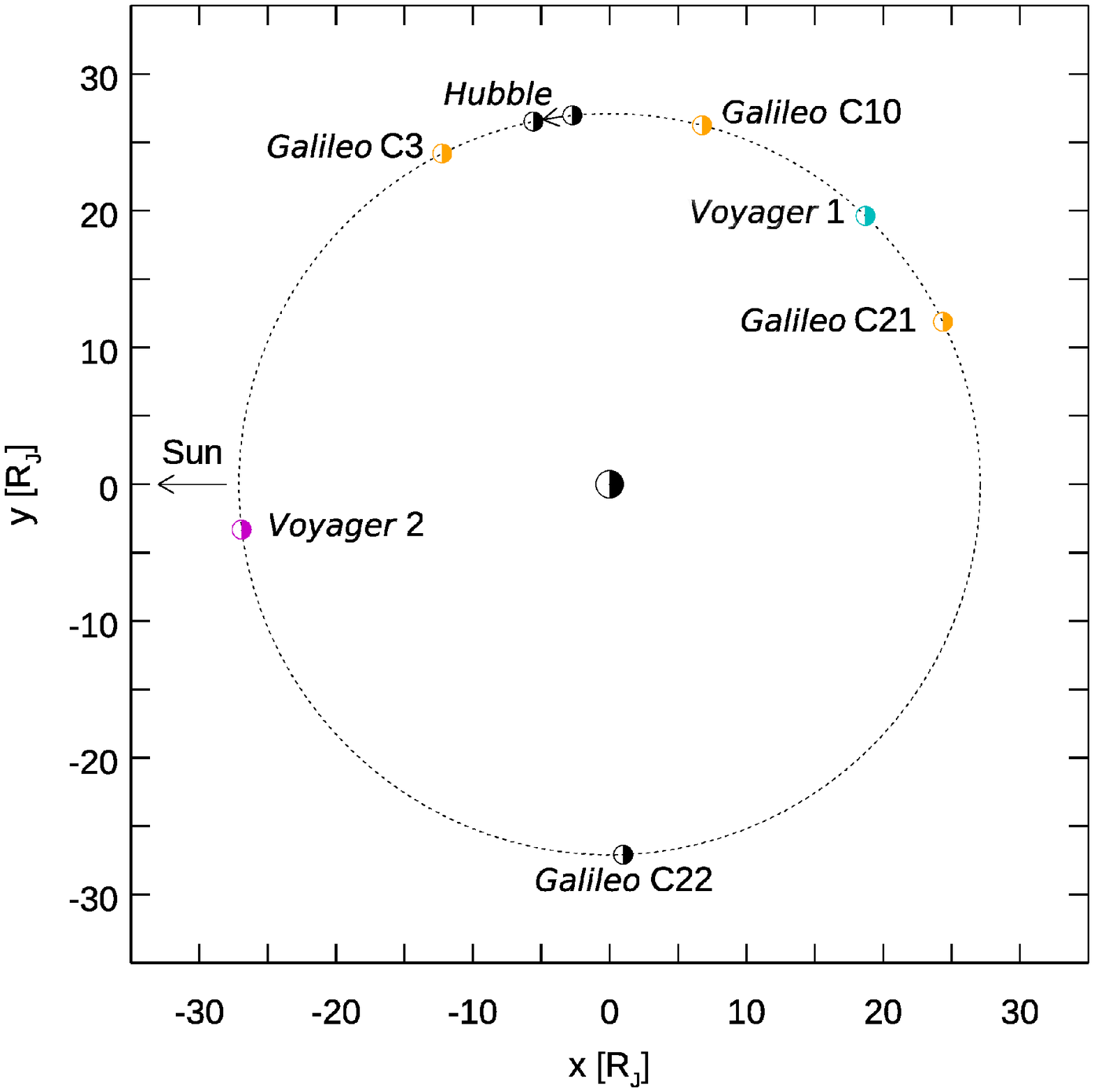}}
    \caption{(a) Effective surface temperatures as a function of local time measured by \textit{Voyager}~1 and 2 (\citealt{hanel1979}, Fig. 7 therein; \citealt{spencer1987c}, Fig. 17 therein) and \textit{Galileo} (\citealt{moore2004}, Fig. 17.7 therein) are plotted as cyan squares, magenta diamonds, and orange circles, respectively. Note the \textit{Galileo} measurements comprise the C3, C10, and C21 encounters. During the \textit{Voyager} 1 and 2 flybys, Callisto's distance from the Sun was 5.29 AU and 5.33 AU, respectively; and the \textit{Galileo} measurements are adjusted to a heliocentric distance of 5.2 AU \citep{moore2004}. Also plotted are the surface temperature distributions used in this study as a function of subsolar latitude (SSL) for $T_\mathrm{dark}$ and $T_\mathrm{ice}$ in red and blue on the day-side (SSL $<$ 90$^\circ$), respectively, as well as the assumed constant night-side temperature, $T_\mathrm{night-side}$, in black. Whereas $T_\mathrm{dark}$(SSL) is calculated assuming radiative equilibrium with Callisto being 5.2 AU from the Sun (Eq. \ref{eq:T_dark}), $T_\mathrm{ice}$(SSL) is calculated according to the disk-averaged H$_2$O ice temperature of 115 K derived by \cite{grundy1999} (Eqs. \ref{eq:T_ice_1} -- \ref{eq:T_ice_2}). (b) The orbital configurations of Callisto during the aforementioned \textit{Voyager} and \textit{Galileo} flybys in Jupiter radii, $R_J$~=~71,492~km. The empty and shaded halves of the circles representing Callisto and Jupiter represent the illuminated and night-side hemispheres, respectively, with the same colors as the corresponding data points from (a). The dotted line represents Callisto's orbit about Jupiter. We also include the C22 flyby for reference to results discussed in Section \ref{discussion_gurnett} as well as two points representing the start and stop time for the original Hubble observation near eastern elongation in which the H corona was later detected \citep{strobel2002}.}
    \label{fig:T0_vs_LocalTime}
\end{figure}

The relatively warm surface temperatures observed by \textit{Voyager} 1 and 2 \citep{hanel1979, spencer1987c} and by \textit{Galileo} \citep{moore2004} displayed in Fig.~\ref{fig:T0_vs_LocalTime} are consistent with Callisto's predominantly dark surface. If, however, this warm, dark material were ice-enriched (e.g., \citealt{clark1980}), then a substantial amount of H$_2$O would sublimate and preferentially migrate away from the equatorial regions and deposit onto mid- to high-latitude regions in a relatively short amount of time (e.g., \citealt{sieveka1982, spencer1987b}). However, there is no visual evidence for a poleward migration of ice and concomitant polar ice caps \citep{spencer1984, spencer1987b}, suggesting that Callisto has a relatively stable surface, dominated by dark non-ice or ice-poor material with relatively sparse bright, ice patches from which the sublimation rates are low \citep{squyres1980, sieveka1982, spencer1984, spencer1987b}.

Using the temperature dependence of the near infrared water ice reflectance spectrum, \cite{grundy1999} derived much colder disk-averaged H$_2$O ice temperatures at Callisto of $T_\mathrm{ice} \sim 115 \pm$20~K. These temperatures are indicative of the sparse, bright ice patches, which are segregated from the dark non-ice or ice-poor material (e.g., \citealt{spencer1987b}). Moreover, heat conduction between the higher thermal inertia, bright material and the lower thermal inertia, dark material would be negligible such that the two temperatures remain independent of one another. The proposition of segregated patches of ice and non-ice or ice-poor material has been applied to explain images of Callisto's surface \citep{spencer1984, spencer1987b} as well as geological processes \citep{moore1999}. In all previous modeling studies of Callisto's atmosphere, however, only the notion of an intimate mixture of ice and non-ice has been considered in which the ice sublimates at Callisto's warm day-side temperatures producing a locally relatively dense atmospheric component (e.g., \citealt{liang2005, vorburger2015, hartkorn2017}, CM21). Hence, prior to this study, the correlation of Callisto's surface temperatures to water production via sublimation has not been extensively explored. Therefore, as described below, we consider 2 significantly different versions of H$_2$O production: \\

\noindent (1) ``Intimate Mixture,'' hereafter referred to as ``IM:'' the ice and dark non-ice or ice-poor material are intimately mixed (e.g., \citealt{clark1980, roush1990, calvin1991}), and H$_2$O sublimates at Callisto's relatively warm day-side temperatures (e.g., \citealt{liang2005, vorburger2015, hartkorn2017}, CM21), $T_\mathrm{dark}$ (solid red line in Fig. \ref{fig:T0_vs_LocalTime}); and \\

\noindent (2) ``Segregated Patches,'' hereafter referred to as ``SP:'' the bright ice and dark non-ice or ice-poor material are segregated into patches (e.g., \citealt{spencer1987b}), so that H$_2$O solely sublimates from the former at Callisto's day-side ice temperatures \citep{grundy1999}, $T_\mathrm{ice}$ (solid blue line in Fig. \ref{fig:T0_vs_LocalTime}), and the latter sufficiently inhibits sublimation from any underlying ice.

\subsubsection{Intimate Mixture (IM)} \label{sec:IM}

Assuming the ice and dark non-ice or ice-poor material are intimately mixed, we calculate the day-side surface temperature distribution, $T_\mathrm{dark}$ (red line in Fig. \ref{fig:T0_vs_LocalTime}), assuming radiative equilibrium with Callisto being 5.2 AU from the Sun, the average distance of the Jovian system from the Sun, which was the case during the original observation of Callisto's atmosphere from which the H corona was later detected (\citealt{strobel2002}, Table 1 therein). \cite{spencer1987c} showed that among the icy Galilean satellites this assumption at Callisto's surface aligned best with the observed surface temperature profiles (Fig. 26 therein); and because of its relatively long day and low thermal inertia, midday temperatures observed by \textit{Voyager} are only about 5~K below equilibrium value (magenta diamonds and cyan squares vs. red line in Fig. \ref{fig:T0_vs_LocalTime}). $T_\mathrm{dark}$ is calculated on the day-side (SSL $<$ 90$^\circ$) using the radiative equilibrium equation:

\begin{equation}
    \varepsilon \sigma T_\mathrm{dark}^4(\mathrm{SSL}) = (1-A) F \cos(\mathrm{SSL}),
\label{eq:T_dark}
\end{equation}

\noindent where $\varepsilon$ is the emissivity and, as has been done in the literature at Callisto \citep{purves1980, sieveka1982, spencer1984, spencer1987b, grundy1999, moore1999}, is assumed to be unity; $\sigma$ is the Stefan-Boltzmann constant; $A = 0.13$ is the Bond albedo consistent with the literature \citep{morrison1977, johnson1978, squyres1980, squyres1981, spencer1989, buratti1991}; and $F$ is the solar flux at 5.2 AU. Applying Eq. \ref{eq:T_dark} at SSL = 0$^\circ$ results in a subsolar temperature of $\sim$167 K. To resemble the observed temperature maps of Callisto \citep{hanel1979, spencer1987c, moore2004} reproduced in Fig. \ref{fig:T0_vs_LocalTime}, a minimum temperature of 80 K is enforced at the terminator and the night-side, $T_\mathrm{night-side}$ (solid black lines in Fig. \ref{fig:T0_vs_LocalTime}).

\subsubsection{Segregated Patches (SP)} \label{sec:SP}

Assuming the bright ice and dark non-ice or ice-poor material are segregated, to generate a temperature distribution for the segregated ice patches, $T_\mathrm{ice}$ (blue line in Fig. \ref{fig:T0_vs_LocalTime}), according to the disk-averaged ice temperatures derived by \cite{grundy1999} we assume a constant $A$ and unit emissivity and ignore any longitudinal variations so that the radiative equilibrium equation can be integrated across the day-side hemisphere:

\begin{equation}
    \int_0^{\pi / 2} \bigg\{ \sigma T_\mathrm{ice}^4(\mathrm{SSL}) = (1-A) F \cos(\mathrm{SSL}) \bigg\} \sin(\mathrm{SSL}) d\mathrm{SSL} \rightarrow \sigma < T_\mathrm{ice}^4> = \frac{1}{2} (1-A) F,
\label{eq:T_ice_1}
\end{equation}

\noindent where $<>$ represents the hemispheric average. A hemispheric average temperature for the ice can then be calculated as $T_{\mathrm{ice}, \mathrm{avg}} = <T_\mathrm{ice}>^{1/4} = \left( \frac{(1-A) F}{2 \sigma} \right)^{1/4}$. At SSL = 0$^\circ$, the subsolar (SS) ice temperature, $T_{\mathrm{ice}, \mathrm{SS}} = \frac{(1-A) F}{\sigma} = 2^{1/4} T_{\mathrm{ice}, \mathrm{avg}}$, and we apply the disk-averaged temperature of ice at Callisto derived by \cite{grundy1999} for $T_{\mathrm{ice}, \mathrm{avg}}$, 115 K, such that $T_{\mathrm{ice}, \mathrm{SS}} \sim 137$~K. Note we just use one value for $T_{\mathrm{ice}, \mathrm{avg}}$, thereby neglecting the small differences in $T_{\mathrm{ice}, \mathrm{avg}}$ \cite{grundy1999} calculated between LH and TH and do not consider the lower and upper bounds given by the error bars ($\pm \sim$20 K). To calculate a day-side surface temperature distribution for the ice patches we use the following equation:

\begin{equation}
    T_\mathrm{ice} (\mathrm{SSL}) = (T_{\mathrm{ice}, \mathrm{SS}} - T_\mathrm{night-side}) \left( \cos(\mathrm{SSL}) \right)^{1/4} + T_\mathrm{night-side},
\label{eq:T_ice_2}
\end{equation}

\noindent where, like the surface temperature distribution for $T_\mathrm{dark}$, $T_\mathrm{night-side} = 80$ K is set as a constant at the terminator and the night-side (solid black lines in Fig. \ref{fig:T0_vs_LocalTime}). The surface temperature distribution for the dark non-ice or ice-poor patches in this scenario is the same as that calculated above for $T_\mathrm{dark}$ via Eq. \ref{eq:T_dark}. 

\subsection{Sources and Sinks} \label{sources_and_sinks}

\subsubsection{Sublimation and Thermal Desorption} \label{sublimation}

For the IM scenario, i.e., sublimation from ``dirty ice,'' the spatial distribution of the H$_2$O sublimation flux corresponding to $T_\mathrm{dark}$(SSL) is reduced by a factor $f$. Here $f$ is a fit parameter used to constrain H$_2$O production and hence its contribution to the observed H corona. In CM21, $f$ was referred to as an ``ice concentration,'' and a value of $f = 10^{-1}$ was used. Since the subsolar temperature applied here, $\sim$167 K, is roughly 12 K higher than that in CM21, 155 K, we consider a range for $f = 10^{-4} - 10^{-2}$, with the upper bound yielding similar day-side sublimation fluxes and H$_2$O densities as CM21.

For the SP scenario, when simulating sublimation from segregated ice patches, $f$ represents the surficial coverage of ice, for which we use a conservative estimate from the literature of 10$\%$ \citep{spencer1987a}. Although the actual size and location of these patches can certainly have an effect on the corresponding sublimation flux and distribution, due to the lack of locally accurate resolution for such parameters, we simply assume a random distribution for the ice throughout the surface where H$_2$O molecules can sublimate from and return to.

\begin{figure}[t!]
    \centering
    \includegraphics[width=\textwidth]{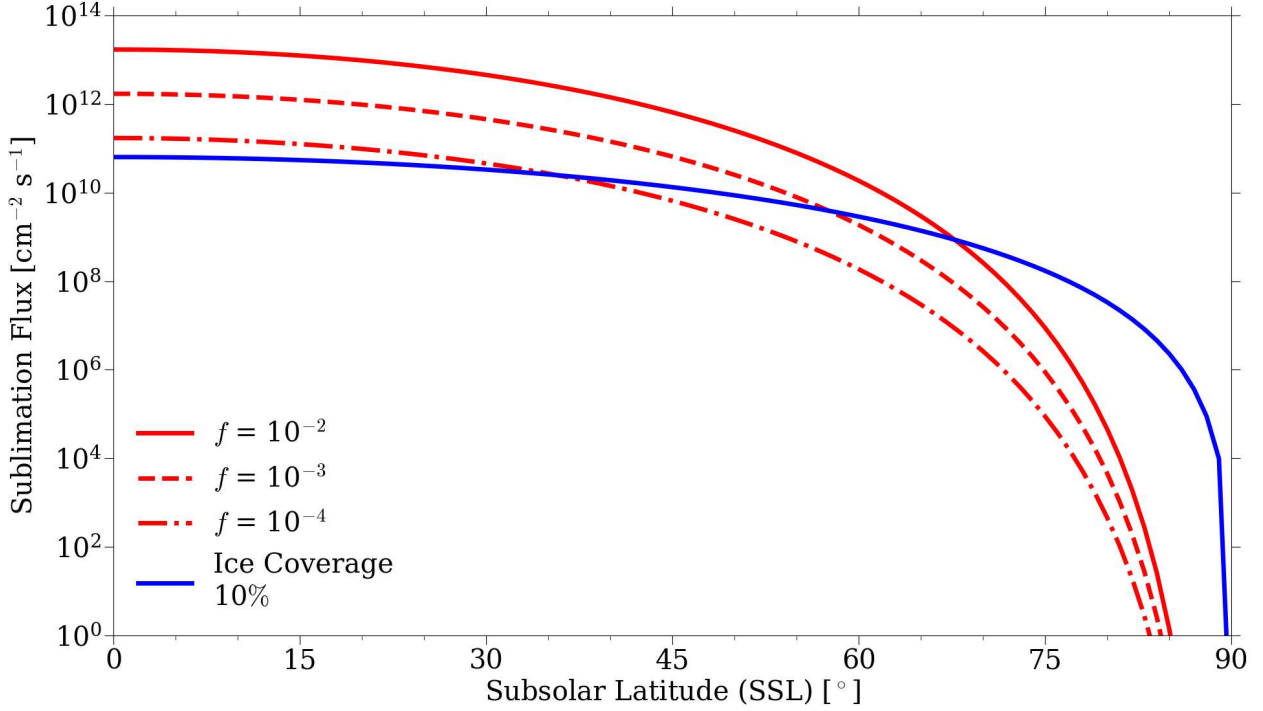}
    \caption{Sublimation fluxes across Callisto's surface for the Intimate Mixture (red lines) and Segregated Patches (blue line) scenarios. The solid, dashed, and dash-dotted red lines represent sublimation fluxes for the Intimate Mixture scenario with a sublimation reduction factor of $10^{-2}$, $10^{-3}$, and $10^{-4}$, respectively; and the solid blue line represents the sublimation flux for the Segregated Patches scenario with an assumed surficial ice coverage of 10$\%$.}
    \label{fig:SSL_vs_subflux}
\end{figure}

The vapor pressure, $P_v$, is calculated using the formula from \cite{feistel2007}:

\begin{equation}
\ln \left( \frac{P_v(T_0(\mathrm{SSL}))}{P_t} \right) = \frac{3}{2} \ln \left( \frac{T_0(\mathrm{SSL})}{T_t} \right) + \left( 1 - \frac{T_t}{T_0(\mathrm{SSL})} \right) \eta \left( \frac{T_0(\mathrm{SSL})}{T_t} \right),
\end{equation}

\noindent where $P_t$ = 6.1166$ \times 10^{-3}$ bar and $T_t$ = 273 K are the triple point pressure and temperature for water, respectively; and $\eta$ is a polynomial relation between the surface temperature, $T_0$, and $T_t$. Depending on the sublimation scenario, $T_0$ can either represent $T_\mathrm{dark}(\mathrm{SSL})$ (IM) or $T_\mathrm{ice}(\mathrm{SSL})$ (SP) illustrated in Fig. \ref{fig:T0_vs_LocalTime} by red and blue lines, respectively. The corresponding sublimation flux, $\Phi_\mathrm{subl}$, is then calculated via the following equation:

\begin{equation}
    \Phi_\mathrm{subl} = \frac{P_v}{\sqrt{2 \pi k_B T_0 m_\mathrm{H_2O}}} f,
\end{equation}

\noindent where $k_B$ is the Boltzmann constant and $m_\mathrm{H_2O} = 18.02$ amu is the mass of an H$_2$O molecule. The sublimation fluxes corresponding to $T_\mathrm{dark}(\mathrm{SSL})$ (IM) and $T_\mathrm{ice}(\mathrm{SSL})$ (SP) are illustrated in Fig. \ref{fig:SSL_vs_subflux} by solid, dashed, and dash-dotted red lines and a solid blue line, respectively. Note that due to the more gradual decrease in $T_\mathrm{ice}$ across the day-side hemisphere compared to the relatively sharp drop-off of $T_\mathrm{dark}$ (Fig. \ref{fig:T0_vs_LocalTime}), $\Phi_\mathrm{subl}$ from the former eventually surpasses that of the latter, regardless of $f$, near the terminator (SSL $\gtrsim$ 75$^\circ$).

Here we treat the radiolytic H$_2$ and O$_2$ components as described in CM20 and CM21: they are assumed to be in steady-state and desorb from the surface with fluxes relative to the local $T_0$. The surface fluxes, $\Phi_{\mathrm{desorb}, i}$, for these species, $i$, are calculated as $\Phi_{\mathrm{desorb}, i} = \frac{1}{4} n_{0, i} v_i$, where $n_{0, i}$ is a prescribed surface density and $v_i = \sqrt{\frac{8 k_B T_0}{\pi m_i}}$ is the Maxwellian speed for each species. For O$_2$, consistent with values suggested by \cite{cunningham2015}, we implement $n_{0, \mathrm{O_2}} \sim$ 10$^9$ cm$^{-3}$; and for H$_2$ we consider a range of values from $n_{0, \mathrm{H_2}} \sim$ (0.4--1)$\times$10$^8$ cm$^{-3}$, as discussed further below.

Finally, sublimating H$_2$O and radiolytically produced H$_2$ and O$_2$ molecules are injected into the domain from cells along the surface using a cosine distribution with velocities sampled from a Maxwellian flux distribution according to the local $T_0$ (e.g., \citealt{brinkmann1970, smith1978}). As discussed in CM20 and CM21, we assume the regolith is permeated with returning H$_2$ and O$_2$; and in the SP scenario, they thermally desorb from both the ice \textit{and} non-ice or ice-poor patches according to the the corresponding local surface temperature, $T_\mathrm{ice}$ or $T_\mathrm{dark}$.

\subsubsection{Photochemical and Electron Impact-Induced Reactions} \label{sect:reactions}

Interactions with photons and magnetospheric electrons ionize and dissociate H$_2$O and H$_2$ producing hot H atoms. Because we are focused on tracking the nascent H as a means to reproduce the observed H corona \citep{roth2017a}, we do not track the other minor products (e.g., O, OH) nor do we consider photochemical and electron impact-induced reactions with O$_2$. We also do not consider ionization of H in the atmosphere as our simulations showed they either return to the surface or they escape the atmosphere much faster than they would be ionized via interactions with photons or magnetospheric electrons.

When an H is created, the excess energy of the reaction is distributed between it and the other products conserving energy and momentum. Since these reactions are rare, the density of H is small compared to that of the other thermal species. Therefore, to improve statistics, each time a reaction occurs that produces H from H$_2$O or H$_2$, we create 100 H particles, each with a weight 1/100 that of its parent species, and scatter them in random directions.

Table \ref{tab:reactions} in Appendix \ref{app:reactions} lists the various reactions, their rates, and the excess energies used in our simulations. To estimate the effect of the magnetospheric electrons we use the number density, $n_e$, and temperature, $k_B T_e$, derived from Voyager measurements by \cite{neubauer1998}: $n_e = 1.1$ cm$^{-3}$ and $k_B T_e = 100$~eV. We assume they can penetrate the extended region of the atmosphere to get a rough upper limit on their effect. However, the local plasma at Callisto is highly variable (e.g., \citealt{galli2022} and references therein) and both temperature and density are not well constrained. Nevertheless, using, for example, the range of smaller electron densities from \cite{kivelson2004} would not affect our principal conclusions. 

Ignoring any contribution from ionization plus recombination in the ionosphere, we use the excess energies producing hot H by photon-induced dissociation and ionization from \cite{huebner2015}. The excess energies resulting in hot H produced by electron impacts are much less certain. If electron energies are low, $\lesssim$16.5 eV, it is often assumed that they only excite the lowest dissociation state of H$_2$ and as a result, the energy released is similar to that of the analogous photon-induced dissociation (Reaction 9 in Table \ref{tab:reactions} in Appendix \ref{app:reactions}); for $k_B T_e \sim$ 100 eV, however, the H$_2$ can be highly excited producing lower energy H fragments \citep{tseng2013}. Therefore, following \cite{tseng2013}, we use the same excess energy for the electron impact-induced dissociation as that for the photochemical reaction which produces Lyman-$\alpha$ from H$_2$ (Reaction 10 in Table \ref{tab:reactions} in Appendix \ref{app:reactions}): 0.488~eV. Because there are insufficient data for the other electron impact-induced reactions, we use, as typically is the case (e.g., \citealt{marconi2007, turc2014, leblanc2017}), the excess energies from the analogous photochemical processes.

For each particle over every $\Delta t$ a reaction occurs if a random number between 0 and 1 is less than the total probability of any reaction occurring, $P_\mathrm{react} = 1 - \exp (- \Sigma_i^N k_{\mathrm{react}, i} \Delta t)$, where $k_{\mathrm{react},i}$ is the reaction rate of the $i$th reaction and $N$ is the total number of reactions. The specific reaction is then selected when another random number between 0 and 1 is eventually less than the sequential ratio $P_{\mathrm{react},j} = \Sigma_i^j k_{\mathrm{react},i} / \Sigma_i^N k_{\mathrm{react},i}$, where $j$ represents the first reaction for which the random number is less than $P_{\mathrm{react},j}$. In the cases where we consider electron impact-induced reactions, since electrons can rapidly move up and down the magnetic field lines continuously flowing past Callisto, we assume for simplicity that electron impact-induced reactions can occur uniformly throughout Callisto's atmosphere. However, photochemical reactions cannot occur in Callisto's shadow; i.e., when $\cos(\mathrm{SSL}_p) < 0$ and $r_p \sin(\mathrm{SSL_p}) < R_C$. Therefore, because the induced reactions differ spatially, we consider them separately; e.g., after first moving a particle and determining if it collides with others, we sequentially determine if a photochemical reaction occurs and, if not, if an electron impact-induced reaction occurs.

The cases for which only photochemical production of H occurs are relevant to either effective ionospheric shielding from the electrons (e.g., \citealt{strobel2002}) and/or Callisto being well outside of Jupiter's plasma sheet. Note that even if Callisto's ionosphere is present but is predominantly produced by a dense, near-surface O$_2$ atmosphere (e.g., \citealt{kliore2002}), then the corresponding ionopause would also remain close to (i.e., a few tens of km above) the surface so that an extended (e.g., hundreds to thousands of km) H$_2$ component would not be well shielded from the impinging thermal plasma regardless. Since Callisto's ionosphere has been observed to be transient \citep{kliore2002}, we also consider interactions with magnetospheric electrons uninhibited by any shielding assuming Callisto is in the center of the plasma sheet as an upper limit for the corresponding contribution to the H corona.

The UV emissions detected by \cite{cunningham2015} were interpreted as photoelectron excited emissions (i.e., airglow). \cite{liang2005} also showed photoelectron-induced reactions to be crucial to reproduce the observed electron densities of \cite{kliore2002} while satisfying the observational constraints of \cite{strobel2002}. When simulating the generation of the H corona from the molecular atmosphere we ignored the photoelectrons, which would a have completely different spatial distribution and a much lower average temperature than the magnetospheric electrons. 

\subsubsection{Boundary Conditions} \label{BCs}

All particles that cross Callisto's Hill sphere at $r_\mathrm{max}$ are assumed to escape from the atmosphere and are removed from the simulation. The surface is assumed to be a source for H$_2$O, O$_2$, and H$_2$ as discussed above, and a sink for these species as well as for H. That is, particles returning to the surface are removed from the simulation as they are assumed to permeate the regolith (e.g., CM20, CM21) and/or react therein. However, we treat the returning H$_2$O particles slightly differently depending on the assumed surface composition: IM or SP. In the SP scenario, we consider 2 different boundary conditions as upper and lower limits for the H$_2$O molecules returning to the surface. \\

\noindent (1) ``All Return:'' all particles are removed from the simulation regardless of the material they land on; if they land on an ice patch they are assumed to condense, and if they land on a non-ice or ice-poor patch they are assumed to permeate this porous regolith and do not re-desorb. \\

\noindent (2) ``Ice Return:'' if a random number between 0 and 1 is less than the surficial coverage of ice, $10\%$, then the particle is assumed to have landed on an ice patch and is removed from the simulation; otherwise it is re-emitted at the local $T_\mathrm{dark}$(SSL) after a temperature-dependent residence time, $\Delta t_\mathrm{res}$, as illustrated in Fig. \ref{fig:T0_vs_TempResTime} in Appendix \ref{app:temp_res_time}. 

Since we do not assume that there are any ice patches on which H$_2$O particles can condense in the IM scenario, but instead that the entire surface is a radiation-altered, porous regolith, as in the ``All Return'' case in the SP scenario when H$_2$O particles return to the non-ice or ice-poor patches, all returning H$_2$O particles are assumed to permeate this material and do not re-desorb.

\subsection{Treatment of Hubble Data and Forward Model} \label{forward_model}

To analyze the likelihood of the scenarios considered, we compare the simulation results with the HST/STIS observation at Lyman-$\alpha$ at eastern elongation reported by \cite{roth2017a}. To do so, we generate a slightly modified version of the forward model used in \cite{roth2017a} and \cite{alday2017} to recreate the brightness at Lyman-$\alpha$ observed by HST/STIS. Here we provide a summary of the main characteristics of the forward model, but a more thorough description can be found in those references.

The H number density from the DSMC simulation for each scenario is integrated along the LOS in the 2D grid illustrated in Fig. \ref{fig:LOS_grid} in Appendix \ref{app:grid}. The resultant H LOS column density, $N_\mathrm{H}$, is then converted into a coronal brightness using $I_\mathrm{H}$~=~10$^{-6}$~$\cdot$~$N_\mathrm{H}$~$\cdot$~$g$, where 10$^{-6}$ is a scaling factor for the unit conversion to Rayleigh and $g$ is the photon scattering coefficient (or $g$-factor), which allows for the calculation of the resonant scattering of solar photons by H atoms \citep{chamberlain1987}. Here we set $g = 1.24\times$10$^{-4}$ s$^{-1}$, the value calculated by \cite{roth2017a} at eastern elongation using the line center solar irradiance for the day of the HST/STIS observation. The simulated H coronal brightness is then combined with sunlight reflected off Callisto's surface, the brightness of the interplanetary medium, as well as the airglow of the Earth's geocorona, with the latter two derived from the HST/STIS data \citep{roth2017a}. Thus, we considered all possible backgrounds for the comparison to the HST observation and present only reasonable assumption and fits.

In addition, we consider the absorption of Lyman-$\alpha$ photons by the H$_2$O atmospheric component, which can be non-negligible for scenarios with high water vapor abundances \citep{roth2017b}. The combination of these parameters allows us to generate a 2D Lyman-$\alpha$ brightness image that is finally convolved with the instrument point spread function \citep{krist2011} to take into account the instrumental effects \citep{roth2014b, alday2017}.

To compare the forward model with the observed HST/STIS image, we average the simulated and observed 2D images using radial bins to generate 1D profiles of Lyman-$\alpha$ brightness as a function of radial distance from Callisto's center. Radial averaging implies that the number of averaged pixels within each bin increases with distance from the center. As a result, the bin at the center has the least number of averaged pixels, leading to a large statistical uncertainty (i.e., large error bar) of the HST/STIS profile in this region. Therefore, we set the radius of the bin at the center to 0.15~$R_C$ to improve the statistics, while every radial bin thereafter has a radius of 0.1~$R_C$.

Finally, while the H contribution to the forward model is directly fed from the simulations, the surface albedo at Lyman-$\alpha$, which is required to model the contribution from the reflected sunlight off Callisto's surface, is not known. To estimate it, equivalent to the approach used by \cite{roth2017a}, we use a Levenberg-Marquardt algorithm to minimize the difference between the simulated and measured intensities and fit the surface albedo. However, because the atmospheric signals overlap with surface reflections on the disk, the region off the disk (just above the limb) contains the most reliable data points for constraining the contributions to the H corona. That means the off-disk signal, and in particular the signal at all points more than 3 pixels ($\sim$200 km) away from the limb, is not affected by any uncertainties in modeling the reflectance signal.

\section{Results} \label{results}

We simulate Callisto’s atmosphere in 2D using the DSMC method \citep{bird1994} to examine the roles of sublimated H$_2$O and radiolytically produced H$_2$ as sources of the H corona observed by HST/STIS at eastern elongation \citep{roth2017a} by modeling the nascent H produced by their interactions with photons and magnetospheric electrons. Our goal is to place constraints on the surface source rates and resultant densities for H$_2$O and H$_2$. A summary of the components of and the corresponding assumptions implemented in the simulated atmospheres presented below can be found in Table \ref{results_summary}. Physical parameters of Callisto used in various calculations below are listed in Table \ref{tab:Callisto_params} in Appendix \ref{app:phys_params}.

\begin{table} [ht!]
    \centering
    \caption{Atmospheric components simulated and their assumed sources.}
    \begin{tabularx}{\textwidth}{|c|X|}
        \hline
        \textbf{Component ($a$\tnote{a})} & \textbf{Source} \\
        \hline
        \multirow{2}{*}{H$_2$O} & Sublimation with the ice and non-ice or ice-poor material either intimately mixed ($b$\tnote{b}) or segregated into patches ($b$). \\
        \hline
        \multirow{3}{*}{H$_2$} & Steady-state thermal desorption of radiolytic product accumulated in the regolith ($c$\tnote{c}) with a uniform surface density ranging from $\sim$(0.4--1)$\times$10$^8$~cm$^{-3}$ ($d$\tnote{d}). \\
        \hline
        \multirow{2}{*}{O$_2$} & Steady-state thermal desorption of radiolytic product accumulated in the regolith ($c$) with a uniform surface density of $\sim$10$^9$ cm$^{-3}$. \\
        \hline
        \multirow{2}{*}{H} & Dissociation of H$_2$O ($e$\tnote{e}) and H$_2$ ($f$\tnote{f}) molecules via interactions with photons ($g$\tnote{g}) and magnetospheric electrons ($g$). \\
        \hline
    \end{tabularx}
    \label{tab:torus}
    \begin{tablenotes}[flushleft]\footnotesize
    \item[a] ($a$) Here we simulate H$_2$O+H (Figs. \ref{fig:H2O_RadColDens}--\ref{fig:H_by_H2O_LOSColDens} and \ref{fig:H_corona_singlespecies}), H$_2$+H (Figs. \ref{fig:H2_numdens_LOS_coldens}--\ref{fig:H_corona_singlespecies}), H$_2$O+H$_2$+H (Figs. \ref{fig:H2O_RadColDens}--\ref{fig:H_by_H2O_LOSColDens} and \ref{fig:H_from_H2O_H2}), and H$_2$O+H$_2$+O$_2$+H (Figs. \ref{fig:H_corona_final}--\ref{fig:col_dens_final}) atmospheres, the last model being our most sophisticated, as it includes thermal and non-thermal collisions as well as interactions with photons and magnetospheric electrons.
    \item[b] ($b$) See Section \ref{sec:surf_temp} for the description of these scenarios.
    \item[c] ($c$) See Section \ref{intro} and CM20 and CM21 for a more thorough description.
    \item[d] ($d$) The upper bound and lower bounds correspond to H being produced solely via photodissociation and via photo- and electron impact-induced dissociation, respectively (e.g., Fig. \ref{fig:H_by_H2_LOSColDens}).
    \item[e] ($e$) When H$_2$O is the primary source of H, the H corona observed by HST/STIS \textit{cannot} be reproduced (e.g., Figs. \ref{fig:H_corona_singlespecies}a--c and \ref{fig:H_from_H2O_H2}).
    \item[f] ($f$) When H$_2$ is the primary source of H, the H corona observed by HST/STIS \textit{can} be reproduced (e.g., Figs. \ref{fig:H_corona_singlespecies}c--d and \ref{fig:H_corona_final}a--b). Note minor contributions \textit{off the disk} from reflected sunlight (e.g., Figs. \ref{fig:H_corona_singlespecies}c--d and \ref{fig:H_corona_final}a--b) and the H produced from H$_2$O (e.g., Fig. \ref{fig:H_corona_final}a--b), which is constrained to a peak sublimation flux of $\lesssim$~10$^{12}$~H$_2$O~cm$^{-2}$~s$^{-1}$ and peak number density of $\lesssim$~10$^8$~H$_2$O~cm$^{-3}$, are included (e.g., Figs. \ref{fig:H_corona_singlespecies}d and \ref{fig:H_corona_final}b).
    \item[g] ($g$) See corresponding reaction rates in Table \ref{tab:reactions} in Appendix \ref{app:reactions}.
    \end{tablenotes}
    \label{results_summary}
\end{table}

\subsection{H from H$_2$O} \label{results:H2O}

\begin{figure}[t!]
    \centering
    \includegraphics[width=\textwidth]{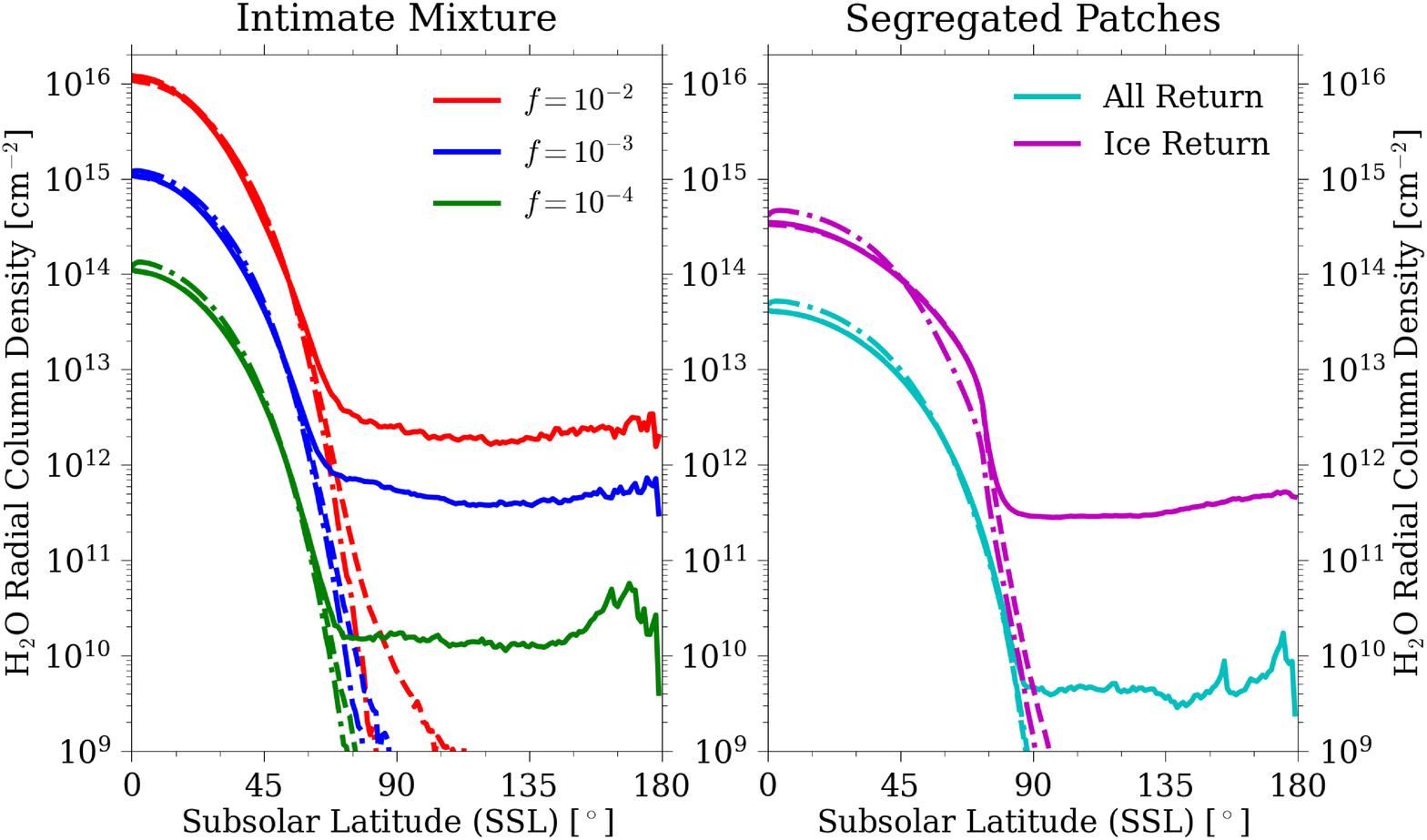}\\
    \includegraphics[width=0.5\textwidth]{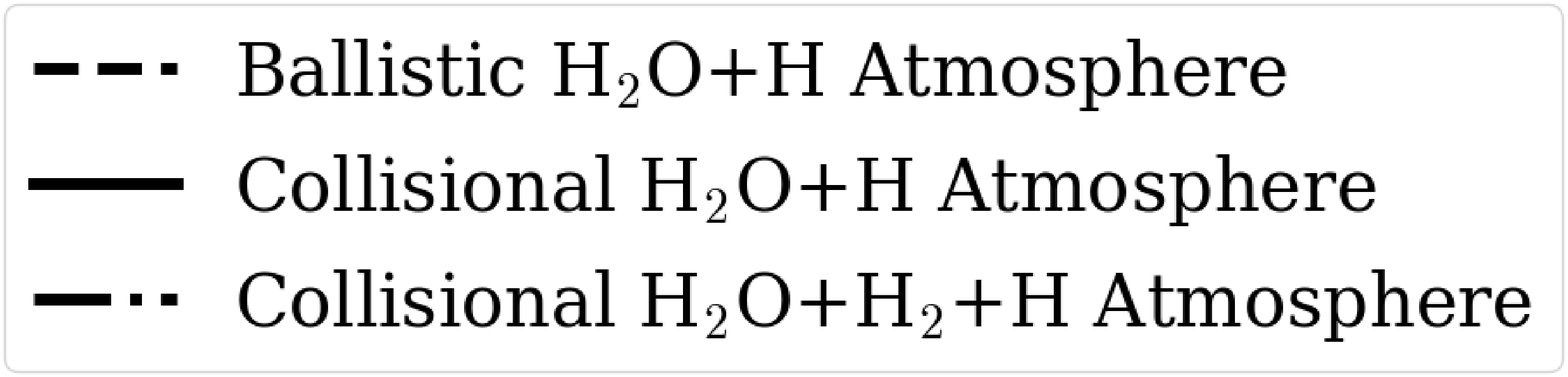}
    \caption{H$_2$O radial column densities generated in the Intimate Mixture and Segregated Patches simulations. $f$ is the sublimation reduction factor implemented in the former (Section \ref{sublimation}), which varies from 10$^{-2}$ (red lines) to 10$^{-3}$ (blue lines) to 10$^{-4}$ (green lines), and ``All Return'' (cyan lines) and ``Ice Return'' (magenta lines) represent the boundary conditions enforced in the latter (Section \ref{BCs}). Solid and dashed lines represent simulations where H$_2$O+H collisions were considered and neglected in an H$_2$O+H atmosphere, respectively, and dash-dotted lines represent atmospheres where collisions with H$_2$ assuming $n_{0, \mathrm{H_2}} \sim 10^8$ cm$^{-3}$ were also considered.}
    \label{fig:H2O_RadColDens}
\end{figure}

For an icy surface whose composition is an Intimate Mixture (IM), sublimation of H$_2$O is determined by an assumed water ice concentration, which we represent with a sublimation reduction factor, $f$, and the local surface temperature, $T_\mathrm{dark}$. Consistent with similar simulations in CM21, by applying the results in Fig. \ref{fig:SSL_vs_subflux} the corresponding H$_2$O atmosphere is most dense near the subsolar point, and drops off several orders of magnitude as one approaches the terminator (Fig. \ref{fig:H2O_RadColDens}). For an icy surface with Segregated Patches (SP) where sublimation of H$_2$O solely occurs from relatively cold ice patches covering only 10$\%$ of the surface at the local ice temperature, $T_\mathrm{ice}$, by applying the result in Fig.~\ref{fig:SSL_vs_subflux} the distributions are similar to those in the IM model, but the peak densities near the subsolar point can be much less, albeit if H$_2$O molecules are able to re-desorb back into the atmosphere, the density can be enhanced roughly an order of magnitude due to the diminished loss rate to the surface (cyan vs. magenta lines in Fig. \ref{fig:H2O_RadColDens}). Since the local temperature in the IM model is determined by the dark material and, as a result, is much warmer than the bright ice patches (e.g., Fig. \ref{fig:T0_vs_LocalTime}: at SSL = 0$^\circ$, $T_\mathrm{dark}$ is $\sim$30 K warmer than $T_\mathrm{ice}$) it is not surprising that it can exhibit orders of magnitude larger densities depending on the ice concentration. In addition, implementing a temperature-dependent residence time, $\Delta t_\mathrm{res}$, effectively halts the H$_2$O migration for SP beyond SSL $\gtrsim$ 75$^\circ$, where $\Delta t_\mathrm{res}$ is on the order of Callisto’s orbital period, $t_\mathrm{orb} \sim$ 1.44$\times$10$^6$~s (Fig. \ref{fig:T0_vs_TempResTime} in Appendix \ref{app:temp_res_time}). That is, there is no further migration beyond this region for returning particles, and as a result, there is a sharp drop-off in density thereafter.

In a single-species H$_2$O atmosphere for either sublimation scenario, consistent with CM21, including or neglecting H$_2$O+H$_2$O collisions has a negligible effect on the density distribution. In an H$_2$O+H atmosphere, however, H$_2$O+H collisions can transfer enough energy to the H$_2$O molecules to reach the night side, thereby populating a night-side H$_2$O atmosphere (Fig. \ref{fig:H2O_RadColDens}), and the loss of H$_2$O slightly increases due to non-thermal escape. However, when a thermal H$_2$ component with $n_{0, \mathrm{H_2}} \sim 10^8$ cm$^{-3}$ is included, H$_2$O+H$_2$ collisions quench the non-thermal H$_2$O produced by collisions with the hot H, diminishing any night-side H$_2$O component and inhibiting any escape. These features can be seen when comparing the solid, dashed, and dashed-dotted lines in Fig. \ref{fig:H2O_RadColDens} as follows. The lines representing ballistic H$_2$O+H atmospheres (dashed lines), collisional H$_2$O+H atmospheres (solid line), and collisional H$_2$O+H$_2$+H (dashed-dotted lines) effectively coincide until close to the terminator. Beyond the terminator the solid lines remain relatively flat as a result of H$_2$O molecules migrating to the night-side and/or escaping after colliding with hot H atoms. Conversely, the dashed and dashed-dotted lines sharply drop off prior to the terminator, the former is because H$_2$O molecules simply follow ballistic trajectories and do not interact with the hot H atoms and the latter is because the H$_2$ component fully quenches the non-thermal energy from the H$_2$O molecules via collisions. When $n_{0, \mathrm{H_2}}$ is reduced to $\sim$4$\times$10$^7$ cm$^{-3}$, the influence of H$_2$ is reduced; e.g., the non-thermal energy transferred via H$_2$O+H collisions is not fully quenched by H$_2$O+H$_2$ collisions, and as a result, H$_2$O is able to migrate to the night-side as well as escape. Similarly, the non-thermal energy transferred via O$_2$+H collisions induces O$_2$ escape.

\begin{figure}[t!]
    \centering
    \includegraphics[width=\textwidth]{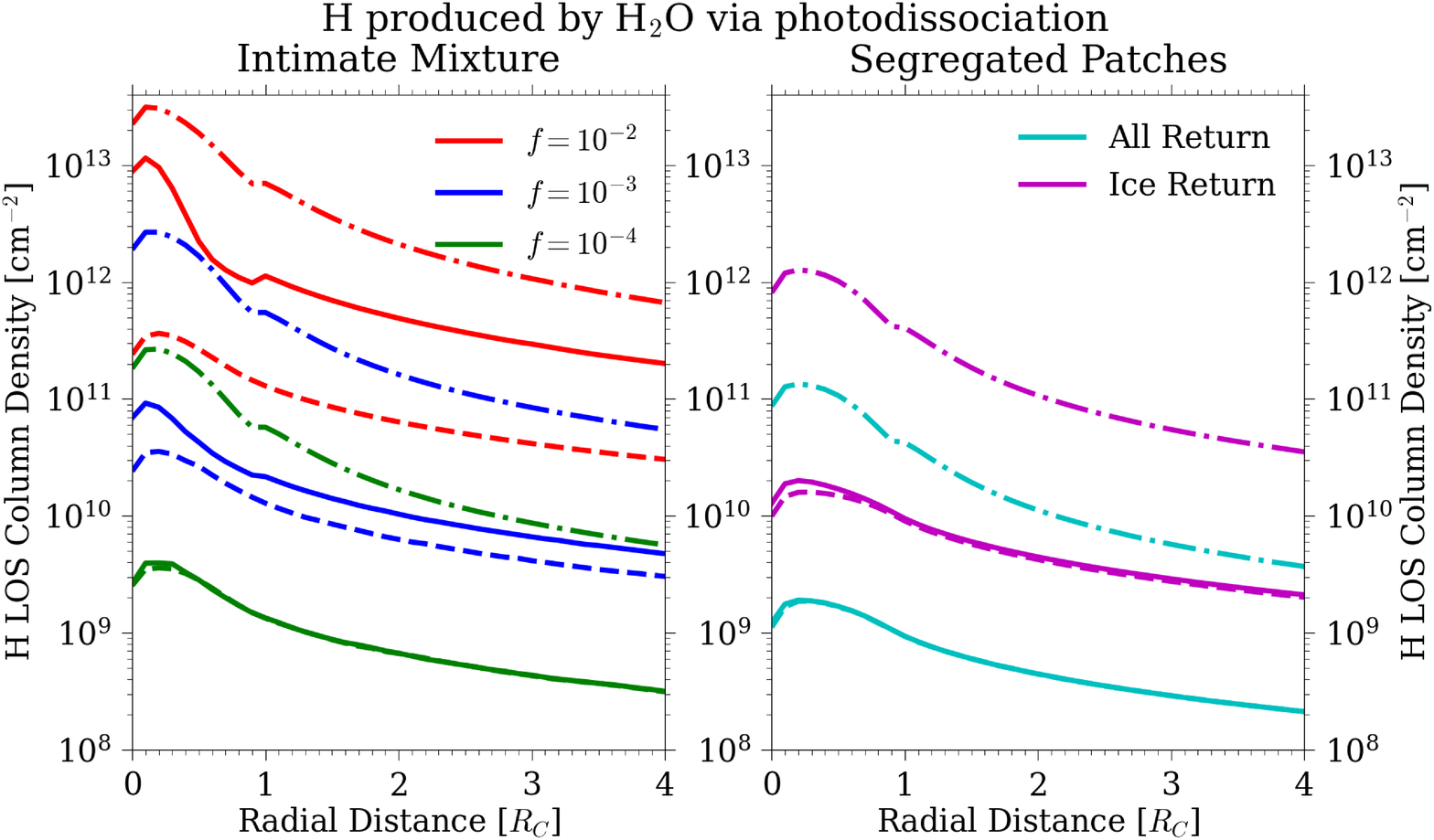}\\
    \includegraphics[width=0.5\textwidth]{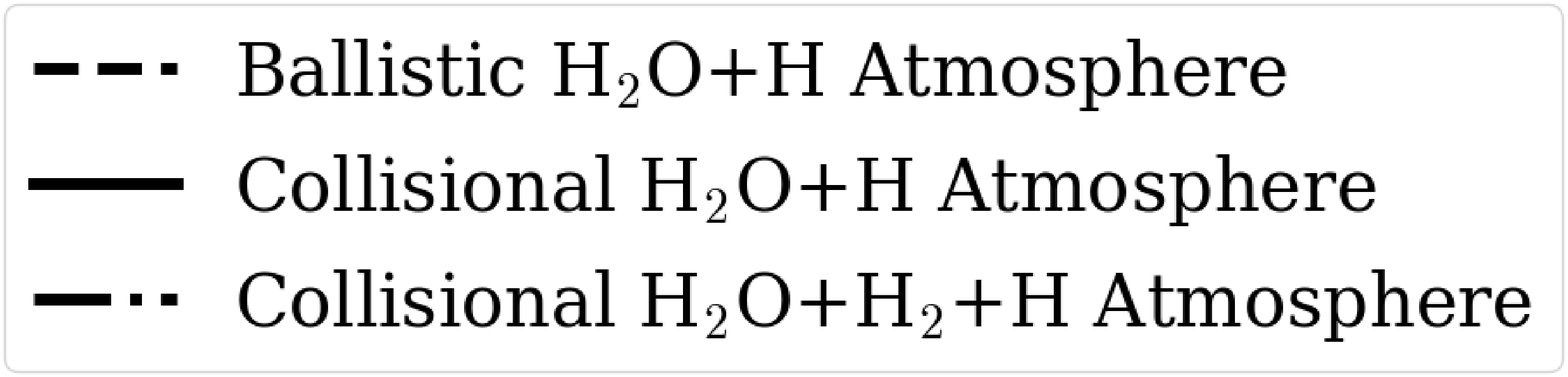}
    \caption{H line-of-sight (LOS) column densities generated in the Intimate Mixture and Segregated Patches simulations. $f$ is the sublimation reduction factor implemented in the former (Section \ref{sublimation}), which varies from 10$^{-2}$ (red lines) to 10$^{-3}$ (blue lines) to 10$^{-4}$ (green lines), and ``All Return'' (cyan lines) and ``Ice Return'' (magenta lines) represent the boundary conditions enforced in the latter (Section \ref{BCs}). Solid and dashed lines represent simulations where H$_2$O+H collisions were considered and neglected in an H$_2$O+H atmosphere, respectively, and dash-dotted lines represent atmospheres where collisions with H$_2$ assuming $n_{0, \mathrm{H_2}} \sim 10^8$~cm$^{-3}$ were also considered. As can be seen, with increasing $f$ (and hence H$_2$O density), the H density increases as it diffuses through this collisional component, and the H density increases even more when the collisional H$_2$ component is included. Note the scale for the LOS column densities are the same as in Fig.~\ref{fig:H_by_H2_LOSColDens} for comparison. The dip in LOS column density at the center of the disk is an artifact of using radial bins for the density simulations and azimuthal bins for the LOS estimate.}
    \label{fig:H_by_H2O_LOSColDens}
\end{figure}

Using the parameters in Table \ref{tab:reactions} in Appendix \ref{app:reactions}, magnetospheric electron-impact dissociation of H$_2$O, producing either one or two H, has a much smaller influence on the net production of H relative to photochemical production. The spatial morphology of H from H$_2$O, regardless of the sublimation scenario, has a similar spatial distribution as its parent: its density peaks near the subsolar point. Therefore, the LOS column density of H also peaks on the disk ($<$1 $R_C$) and drops off rapidly with increasing distance from the disk (Fig. \ref{fig:H_by_H2O_LOSColDens}). Collisions with the H$_2$O serve to slow the H down, thereby reducing its average ``temperature'' (i.e., a measure of the particles' kinetic energy) and enhancing the H density over the disk. Thus, as can be seen in Fig.~\ref{fig:H_by_H2O_LOSColDens}, with increasing $f$ (and hence peak H$_2$O density; e.g., Fig.~\ref{fig:H2O_RadColDens}), the difference between H LOS column densities in collisional (solid lines) and ballistic (dashed lines) H$_2$O+H atmospheres increases, especially when a collisional H$_2$ component is also present (dashed-dotted lines). In a relatively thin H$_2$O atmosphere, however, such as that produced assuming IM with $f=10^{-4}$ or SP, the influence of the H$_2$O+H collisions has a negligible effect on the H component, hence the coinciding solid and dashed green, cyan, and magenta lines in Fig.~\ref{fig:H_by_H2O_LOSColDens}. Also, when collisions with thermal molecules in the atmosphere are neglected, the temperature of the H remains roughly isothermal, and is thus determined by the excess energies of the reactions producing it, resulting in temperatures as high as $\sim$10$^4$ K. However, when collisions with the thermal molecules are taken into account, these temperatures are cooled significantly in the collisional regime of the atmosphere, becoming comparable to those of the thermal molecules; e.g., as low as $\sim$10$^2$ K near the surface.

When H$_2$ and O$_2$ are included, the distribution of H from H$_2$O is significantly affected. With only O$_2$ also present (e.g., CM21) with $n_{0, \mathrm{O_2}} \sim 10^9$ cm$^{-3}$, it can scatter H atoms that would have otherwise rapidly returned to the surface. Such scattering can re-direct H atoms migrating to and descending into the night-side atmosphere upward, producing a secondary peak away from the subsolar point, albeit it is much smaller than the primary peak in the subsolar region where the H is primarily produced. When an H$_2$ component with $n_{0, \mathrm{H_2}} \sim 10^8$~cm$^{-3}$ is also included, not only is the non-thermal energy of the H$_2$O quenched via H$_2$O+H$_2$ collisions, which diminishes its night-side density and inhibits its escape (dash-dotted lines in Fig.~\ref{fig:H2O_RadColDens}), but the density of the H produced from H$_2$O can be further enhanced relative to when only H$_2$O is present by more than an order of magnitude (solid vs. dash-dotted lines in Fig. \ref{fig:H_by_H2O_LOSColDens}). This is due to H$_2$ becoming the dominant collision partner relatively close to the surface, $\lesssim$100 km (e.g., CM21), so that any H produced from H$_2$O has to diffuse through this collisional component en route to returning to the surface or escaping.

\subsection{H from H$_2$} \label{results:H2}

\begin{figure}
    \centering
    \subfloat[]{\includegraphics[width=0.9\textwidth]{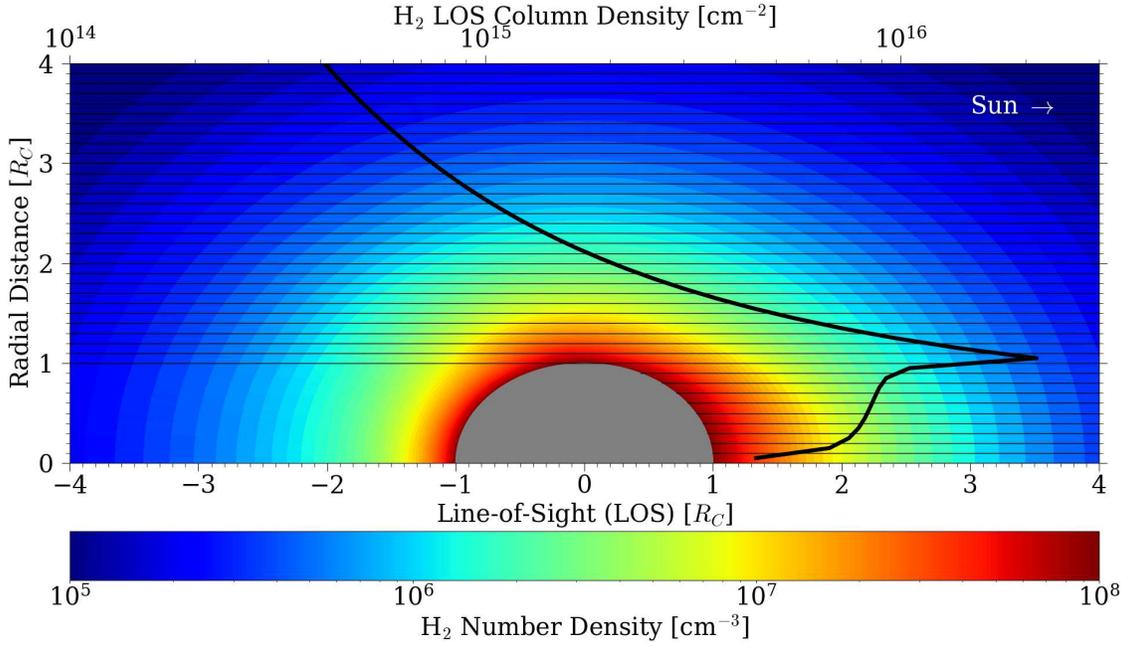}} \\
    \subfloat[]{\includegraphics[width=0.9\textwidth]{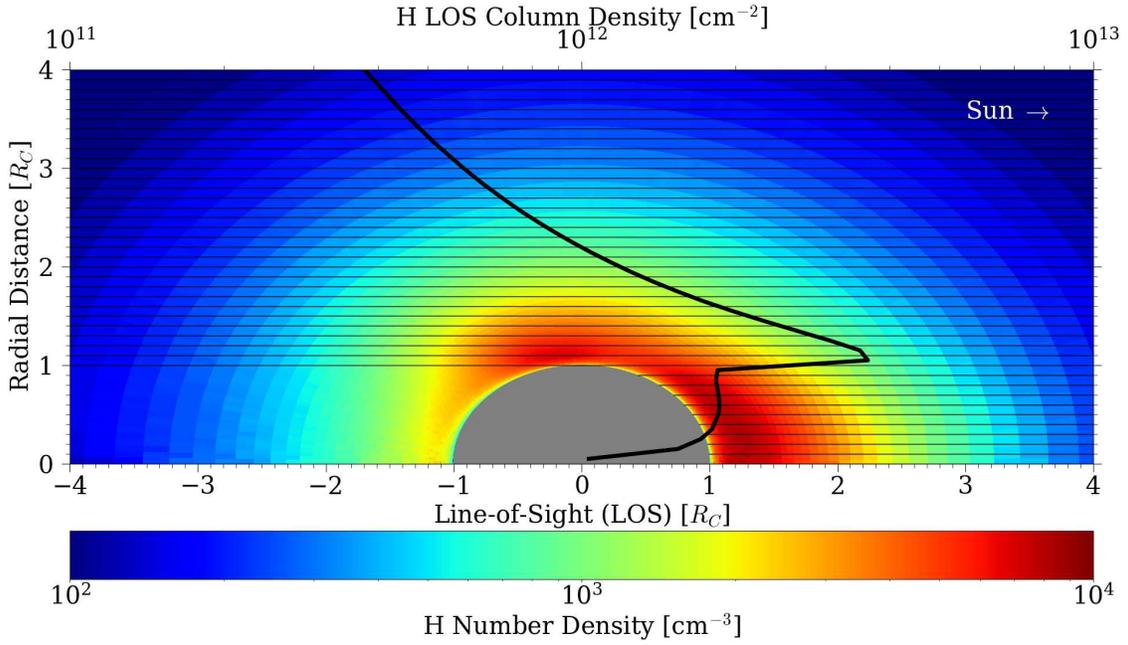}}
    \caption{Number densities (color spectrum) and line-of-sight (LOS) column densities (black line; x-axis, \textit{top}) for (a) radiolytically produced H$_2$ with a surface density of $n_{0, \mathrm{H_2}} \sim 10^8$ cm$^{-3}$ and (b) the H it produces via photochemical reactions in an H$_2$+H atmosphere. The thin black horizontal lines represent the integration bins in which the LOS column densities are calculated with the Sun located to the right of the plot (positive x-direction). The dip in the LOS column densities at the center of the disk is a numerical artifact (see caption of Fig. \ref{fig:H_by_H2O_LOSColDens}).}
    \label{fig:H2_numdens_LOS_coldens}
\end{figure}

A range of steady-state simulations were carried out that only include H$_2$ and H produced by photodissociation and then by photo- and electron impact-induced dissociation. Results are discussed for $n_{0, \mathrm{H_2}} \sim$ 10$^8$~cm$^{-3}$, where only interactions with photons are considered, and for $n_{0, \mathrm{H_2}} \sim$ 4$\times$10$^7$ cm$^{-3}$, the ``intermediate'' case from CM21, where interactions with both photons and magnetospheric electrons are considered. The results from the former are illustrated in Fig. \ref{fig:H2_numdens_LOS_coldens} with an IM surface, which is very similar to that with an SP surface. 

As can be seen in Fig. \ref{fig:H2_numdens_LOS_coldens}, although the distributions of H$_2$ (top panel) and H (bottom panel) are global, both species exhibit density peaks on the day-side. Whereas the relatively small day/night asymmetry for H$_2$ is a result of the diurnal surface temperature gradient, when H is produced only by photo-dissociation the asymmetry is due to the lack of production in Callisto's shadow. For both H$_2$ and H, the sharp peak in the LOS distribution occurs just off Callisto's limb (1--1.1 $R_C$) as expected for a species with a global distribution (e.g., \citealt{roth2017a}). As a result, these simulated LOS profiles are similar to that estimated by \cite{roth2017a} with a slope proportional to the inverse of the radial distance from Callisto squared.

As shown in Fig.~\ref{fig:H_by_H2_LOSColDens}, with the electron-impact induced reaction rates in Table \ref{tab:reactions} in Appendix \ref{app:reactions}, the amount of H$_2$ required to produce the same amount of H is reduced by about 2.5, from $n_{0, \mathrm{H_2}} \sim 10^8$ cm$^{-3}$ (solid red line) to $\sim$0.4$\times$10$^8$ cm$^{-3}$ (solid blue line). Better constraints on the influence of magnetospheric electrons at Callisto's orbit are, of course, needed. Finally, LOS profiles of H were calculated ignoring hot H collisions with H$_2$, as in ballistic calculations of a corona, or including them, as in molecular kinetic models. Although the shape of all of the H LOS column density profiles in Fig. \ref{fig:H_by_H2_LOSColDens} are very similar, it is seen that such collisions can enhance the H component as described in Section \ref{results:H2O} by more than an order of magnitude (red vs. magenta line, blue vs. green lines).

\begin{figure}[ht]
    \centering
    \includegraphics[width=\textwidth]{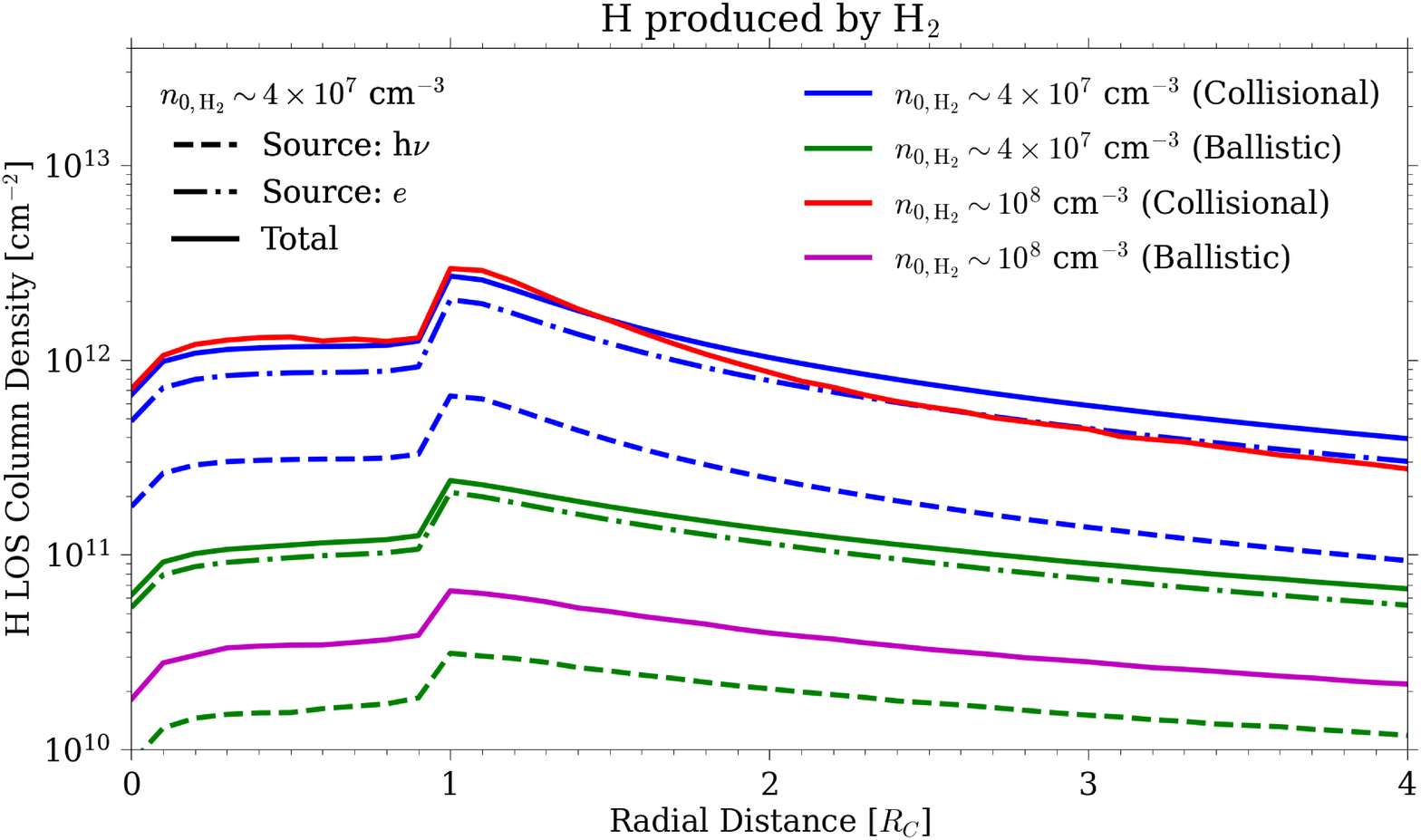}
    \caption{H line-of-sight (LOS) column densities generated in H$_2$+H atmospheres, where H is produced via interactions with photons (h$\nu$) only with $n_{0, \mathrm{H_2}} \sim 10^8$ cm$^{-3}$ (red and magenta lines) and via interactions with h$\nu$ and magnetospheric electrons ($e$) with $n_{0, \mathrm{H_2}} \sim$~4$\times$10$^7$~cm$^{-3}$ (blue and green lines). For the latter, the dashed and dash-dotted lines represent H produced via interactions with h$\nu$ and $e$, respectively, and the solid lines represent the sum total of H. The red solid line and blue lines represent simulations where H$_2$+H collisions were considered and the magenta solid line and green lines represent simulations where they were neglected. Note the scale for the LOS column densities are the same as in Fig. \ref{fig:H_by_H2O_LOSColDens} for comparison. The dip in the LOS column densities at the center of the disk is a numerical artifact (see caption of Fig. \ref{fig:H_by_H2O_LOSColDens}).}
    \label{fig:H_by_H2_LOSColDens}
\end{figure}

Photoabsorption in an H$_2$ component with a surface density of $n_{0, \mathrm{H_2}} \sim$ 10$^8$ cm$^{-3}$ can diminish photodissociation (and hence H production) rates with increasing depth into the atmosphere. We estimated this effect as follows. The H production rate is obtained by first integrating the H$_2$ density along the LOS (x-axis in Fig.~\ref{fig:LOS_grid} in Appendix \ref{app:grid}) obtaining the local LOS column density penetrated as a function of depth, which also varies with radial distance from the LOS (y-axis in Fig.~\ref{fig:LOS_grid} in Appendix \ref{app:grid}), $N$(x, y). The dissociation rate for each reaction, $i$, becomes $\nu_i \exp(-N(\mathrm{x}, \mathrm{y}) \sigma_{\mathrm{abs},i})$, where $\sigma_{\mathrm{abs},i}$ is the relevant photoabsorption cross-section for the processes, $\nu_i$, in Table \ref{tab:reactions} in Appendix \ref{app:reactions}; $\nu_i$ and $\sigma_{\mathrm{abs},i}$ are illustrated as a function of the wavelength in Fig.~\ref{fig:H2_absorbxsec_reactionrates} in Appendix \ref{app:wavelengths}. Fig. \ref{fig:H2_Hproduction_reduction} shows that at most $\sim$13$\%$ of the incoming photon flux is absorbed near the terminator. However, the opacity in this relatively dense H$_2$ atmosphere is even less of an issue when magnetospheric electrons are considered since the H$_2$ density required to reproduce the H corona is reduced (Fig. \ref{fig:H_by_H2_LOSColDens}).

\begin{figure}[ht]
    \centering
    \includegraphics[width=\textwidth]{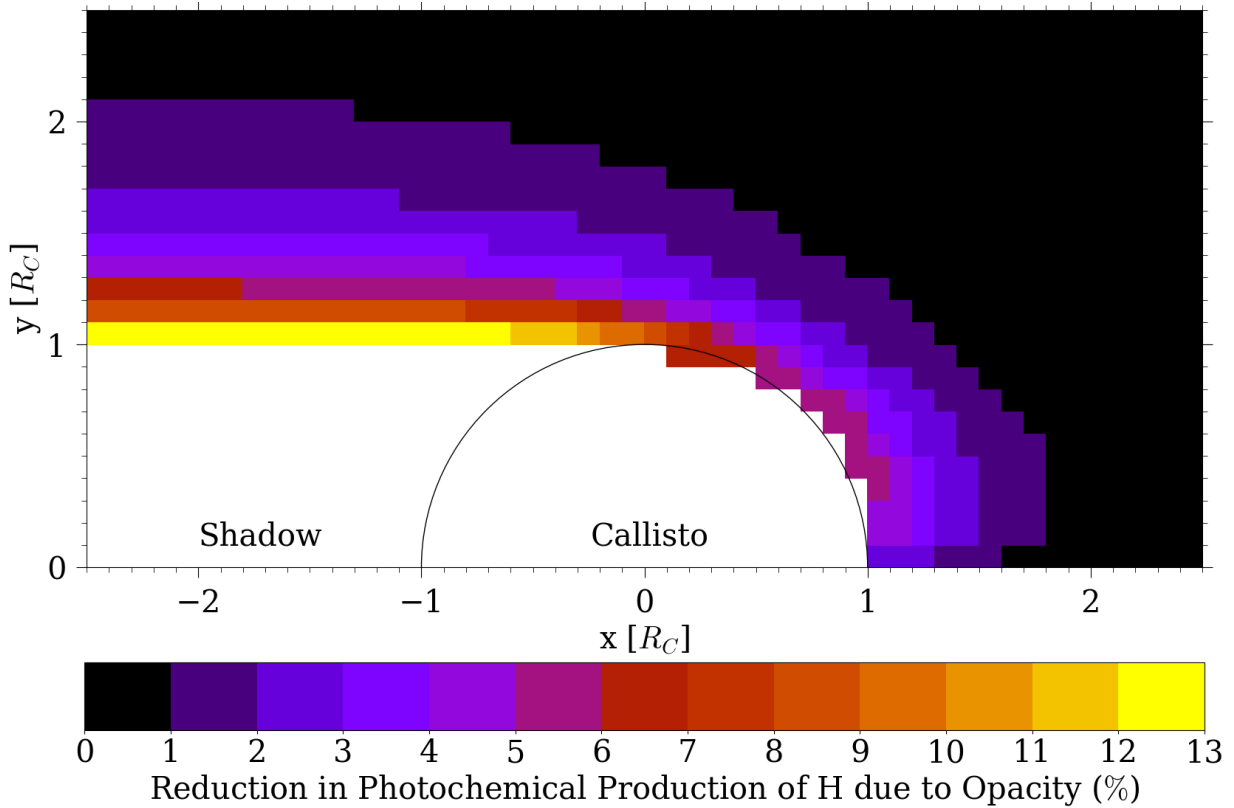}
    \caption{Percent reduction in the photochemical production of H from H$_2$ with a surface density of $n_{0, \mathrm{H_2}} \sim 10^8$ cm$^{-3}$ due to opacity using the wavelength-dependent absorption cross sections and reaction rates illustrated in Fig. \ref{fig:H2_absorbxsec_reactionrates} in Appendix \ref{app:wavelengths}.}
    \label{fig:H2_Hproduction_reduction}
\end{figure}

When collisional H$_2$O and/or O$_2$ components are included they affect the spatial distribution of H$_2$, which in turn affects the spatial distribution of the H produced. Based on our constraints discussed below, H$_2$O only affects H$_2$ in the subsolar region, whereas its interactions with a more dense O$_2$ component dominates globally, albeit the average density of O$_2$ is not well constrained. As both species affect the H$_2$ in the subsolar region, the LOS column densities of the H$_2$ and the H it produces beyond the terminators are only slightly affected. That is, collisions with the heavier thermal species can limit the outflow and migration of H$_2$ on the day-side, which inhibits its migration to and heating and inflation of the night-side atmosphere (e.g., CM21). Since the H distribution primarily follows that of its parent species it is similarly affected. 

\subsection{H Corona: Comparison to Hubble Observation} \label{results:H_corona}

\begin{figure}[t!]
   \centering
    \includegraphics[width=\textwidth]{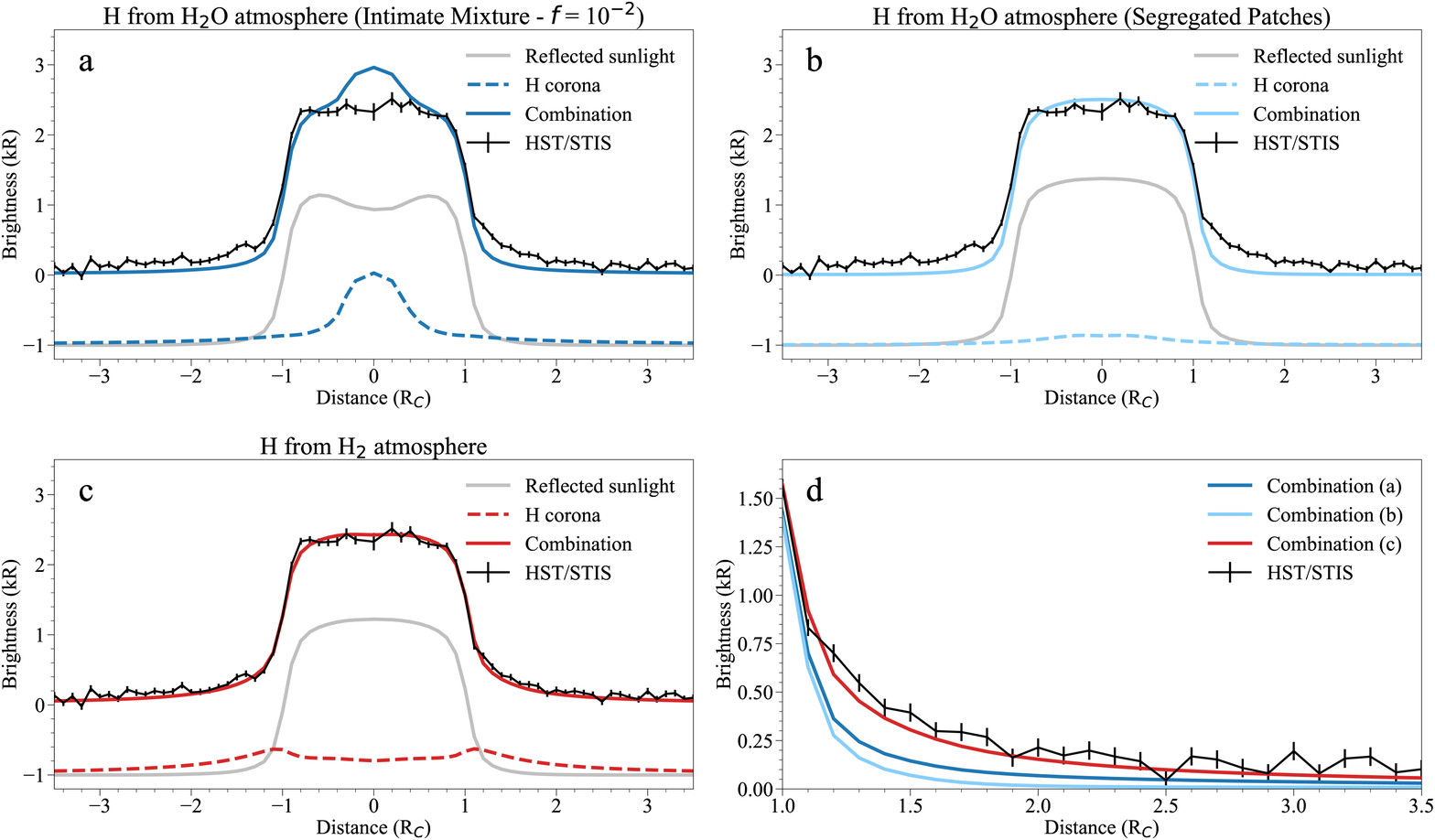}
    \caption{Comparison of the HST/STIS observation with the simulation results for the cases of an H corona produced via photodissociation from H$_2$O in the (a) Intimate Mixture with $f = 10^{-2}$ and (b) Segregated Patches, Ice Return sublimation scenarios, as well as (c) via photodissociation from radiolytically produced H$_2$ with $n_{0, \mathrm{H_2}} \sim$ 10$^8$ cm$^{-3}$. The coronal brightness in (a), (b), and (c) are represented by dark blue, light blue, and red dashed lines, respectively; the black line represents the measured HST/STIS Lyman-$\alpha$ profile with black vertical segments for error-bars of the measurement; the gray line represents the contribution from reflected sunlight; and the forward model results are represented by solid lines of the same color of the coronal brightness. The reflected sunlight in (a) and (b) also includes the absorption at Lyman-$\alpha$ by H$_2$O molecules, hence the dip in at the center of the disk in (a). The reflected sunlight and coronal emissions in (a)--(c) are offset by $-$1~kR to improve the visibility of the figure. Panel d shows a closer look at the forward model results for all cases near the terminator, where the sensitivity of the HST/STIS observation to the H corona is highest. Whereas the axisymmetric simulation results are the same on either side of Callisto's disk, the HST/STIS data presented in (d) are the average of the values from $-1 \rightarrow -3.5 R_C$ and $1 \rightarrow 3.5 R_C$.}
    \label{fig:H_corona_singlespecies}
\end{figure}

For an H$_2$O+H atmosphere using the IM scenario assuming $f$ = 10$^{-2}$, the total modeled corona brightness is seen in Fig. \ref{fig:H_corona_singlespecies}a to exceed the total observed brightness near the subsolar point, where the production of H atoms from H$_2$O peaks. This is the case even when considering the depletion of Lyman-$\alpha$ flux via absorption by the H$_2$O molecules, as indicated by the dip in reflected sunlight in the center of the disk (Fig. \ref{fig:H_corona_singlespecies}a). On the other hand, due to the rapid decrease in sublimation away from the subsolar point, the production of H is too small to replicate the observed brightness beyond the terminator ($>$1 $R_C$; dark blue line in Fig. \ref{fig:H_corona_singlespecies}d), which is also true for the SP models, where there is no depletion of Lyman-$\alpha$ on the disk (Fig.~\ref{fig:H_corona_singlespecies}b) and the production of H is even smaller (light blue line in Fig. \ref{fig:H_corona_singlespecies}d). On the other hand, the distribution of H produced from H$_2$ assuming $n_{0, \mathrm{H_2}} \sim$ 10$^8$ cm$^{-3}$ over the disk, although relatively small relative to the reflected sunlight, is more uniform, increasing from the center of the disk to the terminator until reaching a maximum just off the disk (Fig. \ref{fig:H_corona_singlespecies}c) as in the LOS column density distributions of H$_2$ and H (Figs. \ref{fig:H2_numdens_LOS_coldens}--\ref{fig:H_by_H2_LOSColDens}). The combination of the reflected sunlight and the brightness of the H corona produced from H$_2$ provide a good fit to the HST/STIS measurement (red line in Fig. \ref{fig:H_corona_singlespecies}d).

\begin{figure}[t!]
    \centering
    \includegraphics[width=\textwidth]{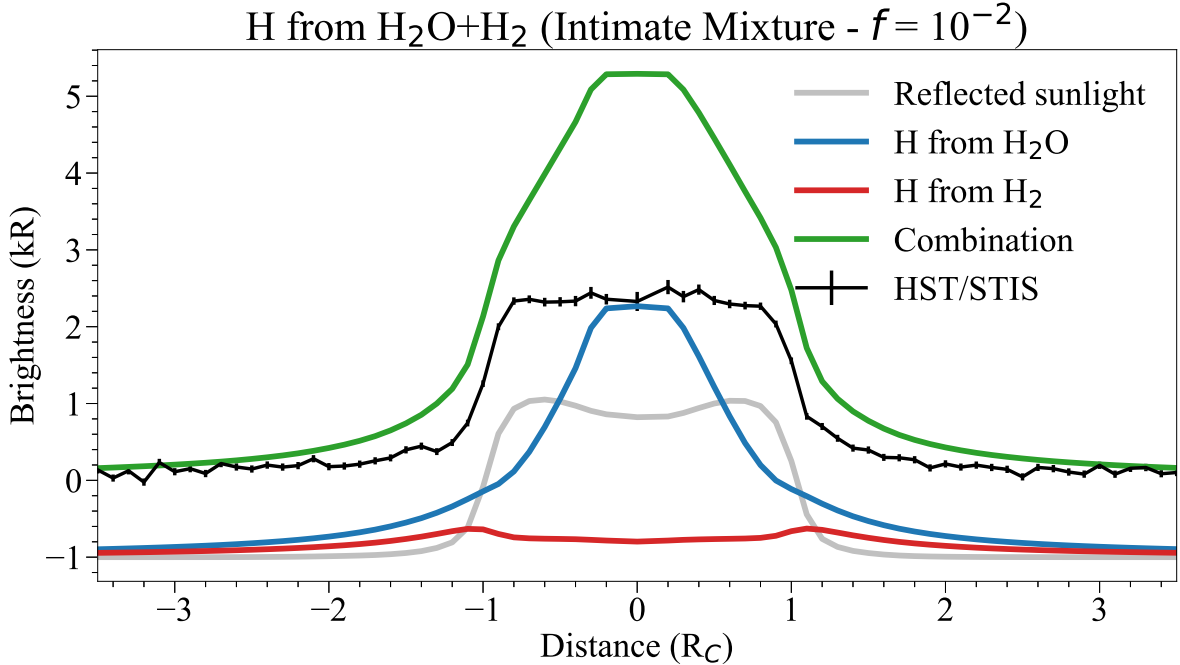}
    \caption{Comparison of the HST/STIS observation with the simulation results for the case of an H corona produced via photodissociation of H$_2$O (dark blue line) and H$_2$ (red line) in an H$_2$O+H$_2$+H atmosphere assuming the Intimate Mixture sublimation scenario with $f$ = 10$^{-2}$ and $n_{0, \mathrm{H_2}} \sim$ 10$^8$ cm$^{-3}$. The black line represents the measured HST/STIS Lyman-$\alpha$ profile with black vertical segments for error-bars of the measurement; the gray line represents the contribution from reflected sunlight including the absorption at Lyman-$\alpha$ by H$_2$O molecules, hence the dip at the center of the disk; and the green line represents results from the forward model. The reflected sunlight and coronal emissions are offset by $-$1~kR to improve the visibility of the figure.}
    \label{fig:H_from_H2O_H2}
\end{figure}

\begin{figure}
   \centering
    \subfloat[]{\includegraphics[width=0.8\textwidth]{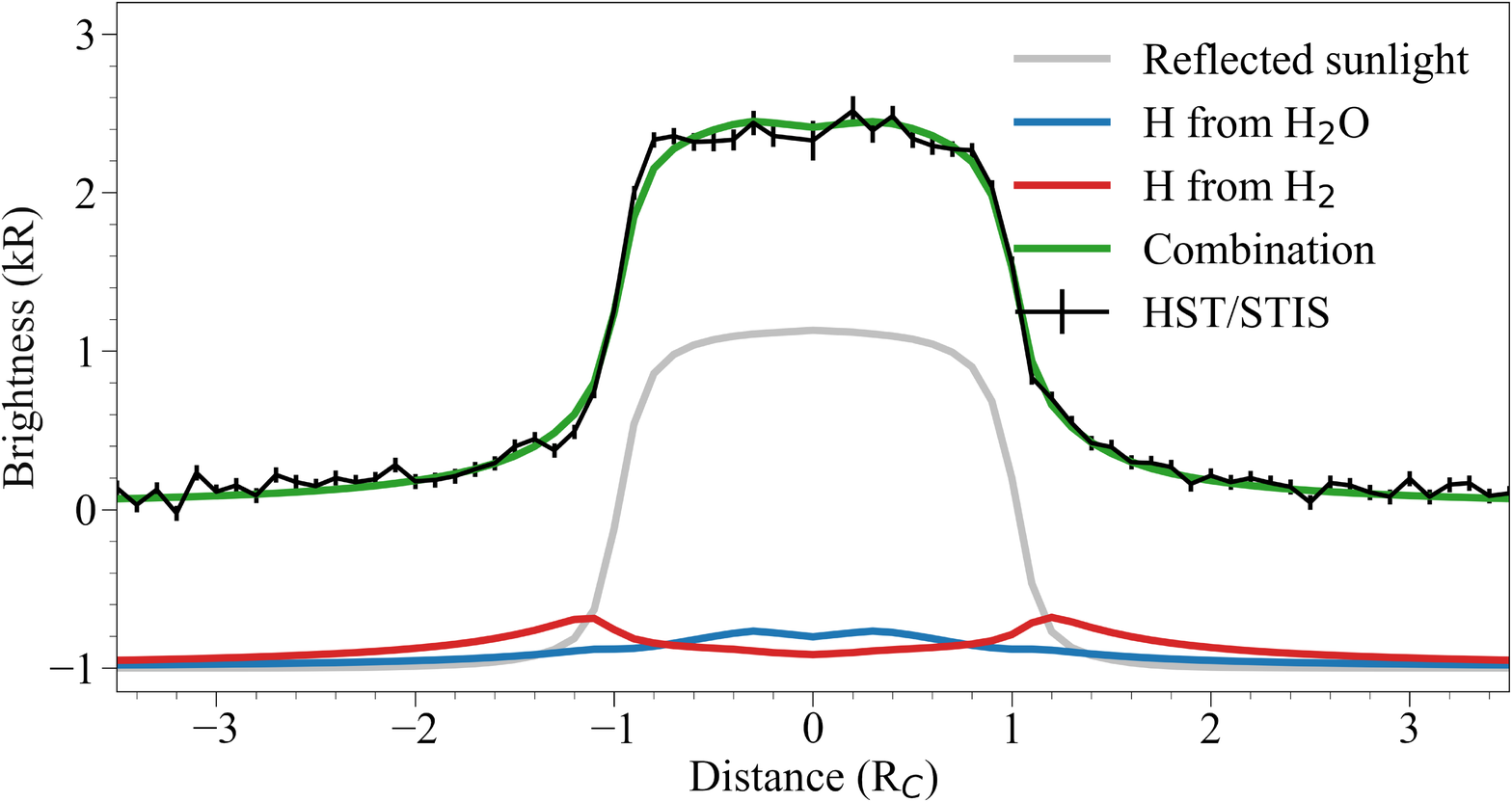}} \\
    \subfloat[]{\includegraphics[width=0.8\textwidth]{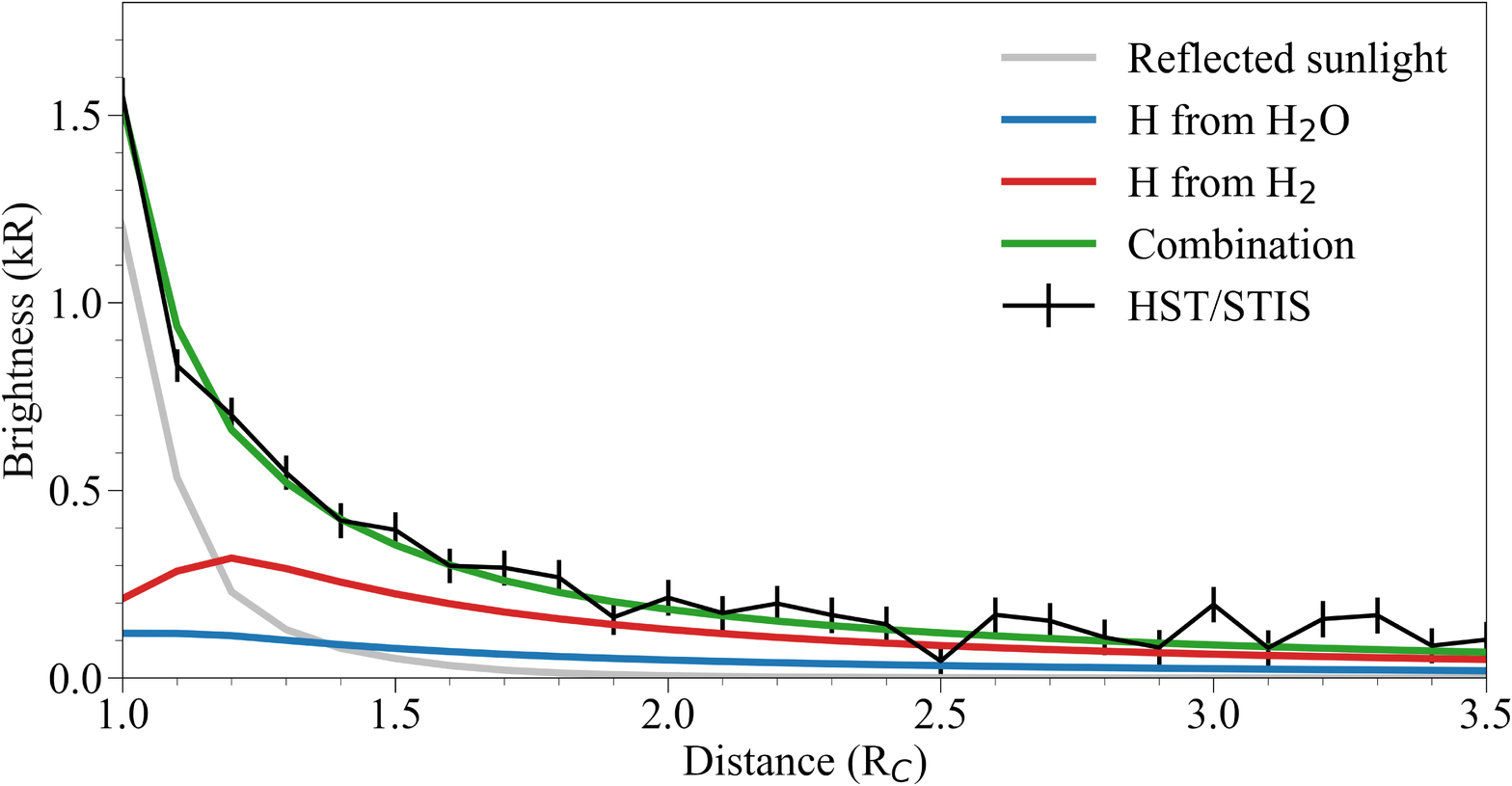}}
    \caption{(a) Comparison of the HST/STIS observation with the simulation results from our most sophisticated model of Callisto's atmosphere: an H$_2$O+O$_2$+H$_2$+H atmosphere subject to interactions with photons and magnetospheric electrons assuming an intimate mixture (IM) surface, a sublimation reduction factor of $f = 10^{-3}$, and O$_2$ and H$_2$ surface densities, $n_0$, of $\sim$10$^9$ cm$^{-3}$ and $\sim$4$\times$10$^7$~cm$^{-3}$, respectively. The H produced from sublimated H$_2$O and radiolytically produced H$_2$ are represented by dark blue and red lines, respectively; the black line represents the measured HST/STIS Lyman-$\alpha$ profile with black vertical segments for error-bars of the measurement; the gray line represents the contribution from reflected sunlight including the absorption at Lyman-$\alpha$ by H$_2$O molecules; and the green line represents results from the forward model. The reflected sunlight and coronal emissions are offset by $-$1~kR to improve the visibility of the figure. (b) A closer look at the results from (a) near the terminator, where the sensitivity of the HST/STIS observation to the H corona is highest. Whereas the axisymmetric simulation results are the same on either side of Callisto's disk, the HST/STIS data presented in (b) are the average of the values from $-1 \rightarrow -3.5 R_C$ and $1 \rightarrow 3.5 R_C$.}
    \label{fig:H_corona_final}
\end{figure}

In addition to scenarios in which H is solely produced from one of the parent species, we also simulated cases including both H$_2$O and H$_2$ as sources of H via interactions with photons and magnetospheric electrons as well as collisions with an O$_2$ component with $n_{0, \mathrm{O_2}} \sim$~10$^9$~cm$^{-3}$. For the IM case a good fit to the data requires $f \lesssim$ 10$^{-3}$ because with a larger $f$ the modeled brightness over the disk becomes even more problematic when collisions with other components are considered (e.g., Fig. \ref{fig:H_from_H2O_H2}), which can create an even larger brightness peak than that in Fig. \ref{fig:H_corona_singlespecies}a and is thus inconsistent with the relatively flat disk profile in the HST/STIS data. However, if $f$ is reduced to improve the agreement over the disk (Fig. \ref{fig:H_corona_final}a), then the H produced from H$_2$O off the disk is $\sim$5$\times$ less than that observed (Fig. \ref{fig:H_corona_final}b). Therefore, the inconsistencies of H$_2$O producing too much H near the subsolar point and too few H away from Callisto's disk relative to the observed H morphology affirm that the H$_2$ component with an average surface density of $n_{0, \mathrm{H_2}} \sim$~4$\times$10$^7$ cm$^{-3}$ is the primary producer of the morphology of the observed H, with minor contributions off the disk from the reflected sunlight and H from H$_2$O (Fig.~\ref{fig:H_corona_final}b). For the SP cases the contribution to the brightness by H produced from H$_2$O is even smaller off the disk (e.g., Fig. \ref{fig:H_corona_singlespecies}d), so the observed data are also predominantly due to surface reflection on the disk and H produced from H$_2$ off the disk. For either case, SP or IM with $f \lesssim 10^{-3}$, although the brightness off the disk is dominated by the H produced from H$_2$, the contributions on the disk are obscured by Lyman-$\alpha$ reflection. That is, any reduction in coronal brightness over the disk does not significantly affect the comparison to the observation, whereas off the disk the comparison is crucial (see Section~\ref{forward_model}).

As described in Section \ref{forward_model}, we assumed the $g$-factor calculated by \cite{roth2017a} when converting the simulated H abundances to the brightness of the scattered sunlight. We estimate an uncertainty of up to 30$\%$, factoring in, for example, the variability of the solar intensity and simplifications about the resonant scattering properties (neglecting all optical thickness effects). Comparing the $g$-factor from \cite{roth2017a} to the scaled Lyman-alpha $g$-factors from \cite{killen2009}, the latter are about 25$\%$ lower (when scaled with distance from the Sun and daily solar flux) confirming our estimated uncertainty. This $\leq$30\% uncertainty translates linearly to the derived abundances everywhere and therefore does not affect our conclusions on the roles of H$_2$O and H$_2$ for producing the observed corona profiles on and off the disk. Although it does affect the derived absolute abundances, the 30$\%$ range is small compared to the uncertainties introduced from the plasma conditions (see Section \ref{sect:reactions}) as well as other effects, such as those discussed later in Section \ref{H2_Discussion}.

As originally suggested in \cite{roth2017a}, a fit to the HST/STIS observation can be generated by a scenario with a nearly uniformly distributed source of H, as is the case in our simulations with H$_2$. Therefore, we considered other possible sources for the observed H. Using the plasma parameters from \cite{vorburger2019} (see Tables \ref{tab:V19_proton}--\ref{tab:CEX} in Appendix \ref{app:phys_params}), proton charge-exchange with all atmospheric species considered in the models as well as with the observed CO$_2$ and O components produces a negligible source rate of H. In addition, \cite{vorburger2015} assumed H was produced via sputtering from hydrated non-ice surface materials; however, the peak number densities they obtained were $\sim$10$^0$~cm$^{-3}$ (Fig. 3 therein). Thus, there is no sufficient \textit{direct} surface source of H that would reproduce the observation. 

\begin{figure}[ht!]
    \centering
    \subfloat[]{\includegraphics[width=0.5\textwidth]{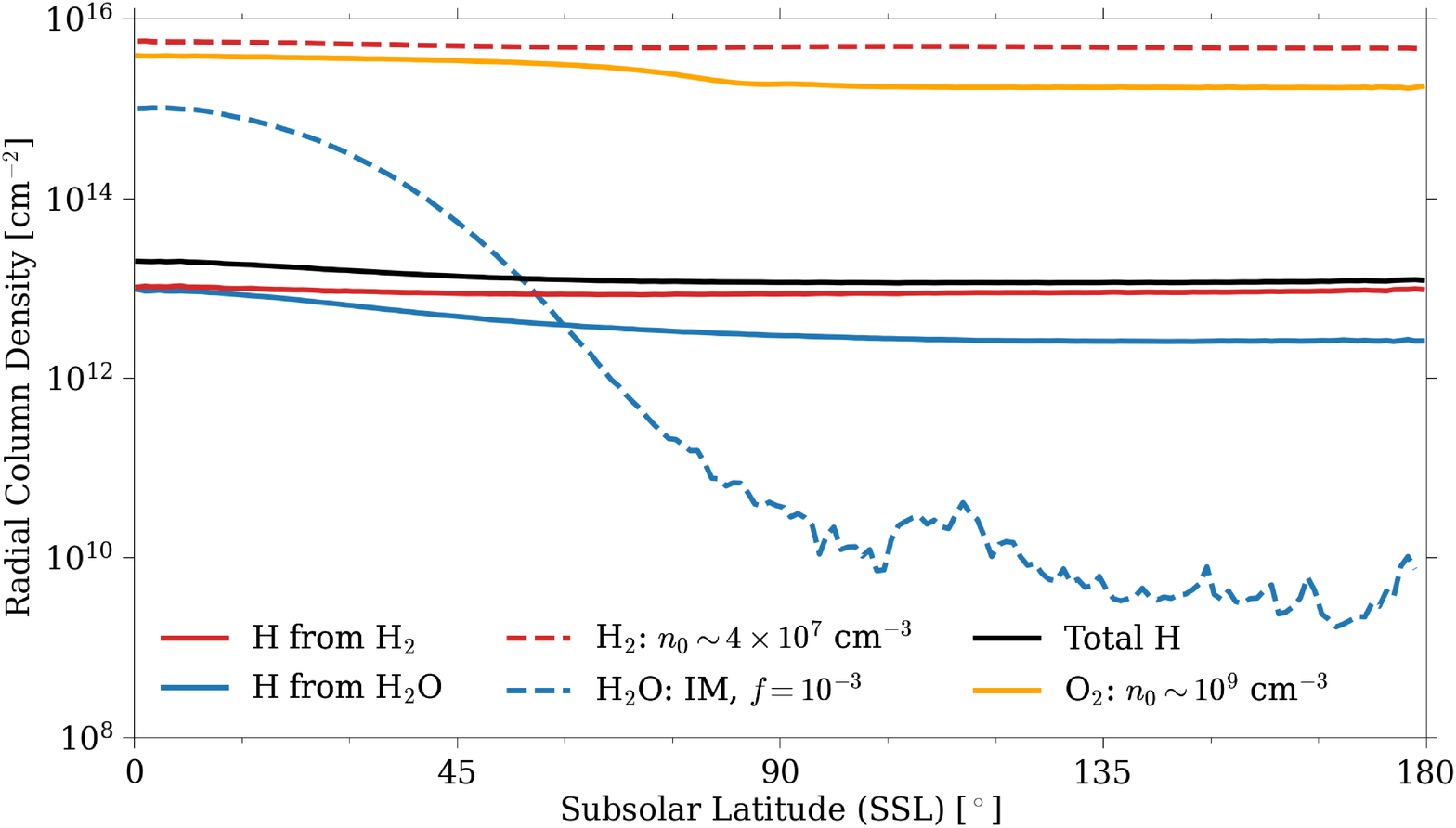}}
    \subfloat[]{\includegraphics[width=0.5\textwidth]{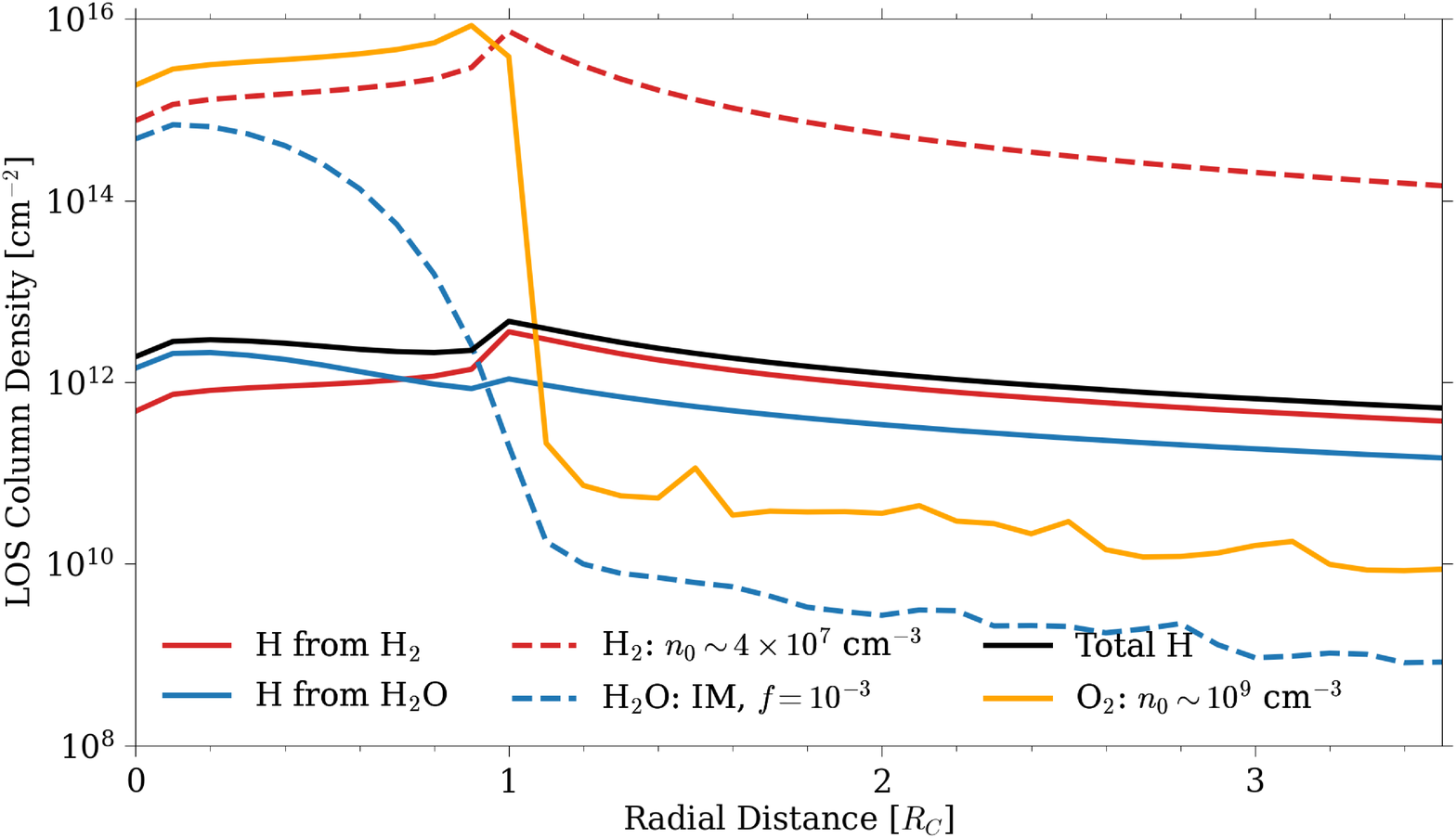}}
    \caption{Results from our most sophisticated model of Callisto's atmosphere: an H$_2$O+O$_2$+H$_2$+H atmosphere subject to interactions with photons and magnetospheric electrons assuming an intimate mixture (IM) surface, a sublimation reduction factor of $f = 10^{-3}$, and O$_2$ and H$_2$ surface densities, $n_0$, of $\sim$10$^9$ cm$^{-3}$ and $\sim$4$\times$10$^7$~cm$^{-3}$, respectively. Radial and line-of-sight (LOS) column density profiles are presented in (a) and (b), respectively, with results for H$_2$O and H$_2$ presented in dashed dark blue and red lines, respectively, the H they produce in the same colors in solid lines, the total H produced in solid black, and O$_2$ in orange. From SSL=0--180$^\circ$ the average Knudsen number at the surface Kn$_{0, \mathrm{avg}} = \frac{\Sigma_i \mathrm{Kn}_{0,i} n_{0,i}}{\Sigma_i n_{0,i}} \sim 0.02-0.05$, where $i$ represents the species, and Kn~=~$\ell / H$ is a dimensionless number used to determine the rarefaction of the atmosphere, with $\ell$ and $H$ the local mean free path and scale height, respectively. With this reduced $n_{0, \mathrm{H_2}}$ the non-thermal energy transferred via H$_2$O+H collisions is not fully quenched by H$_2$O+H$_2$ collisions so that H$_2$O molecules are able to migrate to the night-side (a) and escape, resulting in the thin extended component beyond Callisto's disk shown in (b). Similarly, the non-thermal energy transferred via O$_2$+H collisions induces O$_2$ escape resulting in the thin extended component shown in (b). The dip in the LOS column densities at the center of the disk is a numerical artifact (see caption of Fig. \ref{fig:H_by_H2O_LOSColDens}).}
    \label{fig:col_dens_final}
\end{figure}

In Fig. \ref{fig:H_corona_final}--\ref{fig:col_dens_final} we present results from our most sophisticated model of Callisto's atmosphere composed of H$_2$O, H$_2$, O$_2$, and H subject to interactions with photons and magnetospheric electrons in which the H$_2$ component with an average surface density of $n_{0, \mathrm{H_2}} \sim$~4$\times$10$^7$~cm$^{-3}$ is the primary producer of the morphology of the observed H, the H$_2$O component assuming an IM surface with $f \lesssim$ 10$^{-3}$ is only a minor producer, and the spatial distributions of the atmospheric components are affected by thermal and non-thermal collisions. Our simulation results are dependent on the input parameters we implement, such as the assumed dissociative reaction rates which produce H. However, for the reasons described above, our principal conclusions that H$_2$ is the primary source of Callisto's H corona and that H$_2$O cannot be the primary source are not affected by varying such parameters.

\section{Discussion and Constraints} \label{discussion}

Here we discuss the implications of our results and the rough constraints on sublimation of H$_2$O and the amount of H$_2$ produced primarily from radiolysis.

\subsection{Sublimated H$_2$O}

When the ice on Callisto's surface is intimately mixed with the dark non-ice or ice-poor material and H$_2$O is assumed to desorb from Callisto's warm, day-side surface at a relatively small fraction of the ice sublimation rate ($f = 10^{-2}$), more than enough H is produced to account for the HST/STIS observation of a H corona at Callisto. However, the spatial distribution of H is not consistent with the observation. That is, the LOS column density of the simulated H peaks near the subsolar point, which is inconsistent with what is seen on the disk by HST/STIS (e.g., Figs.~\ref{fig:H_corona_singlespecies}a~and~\ref{fig:H_from_H2O_H2}), and too few H is produced off the disk (e.g., Fig. \ref{fig:H_corona_singlespecies}d). Moreover, relatively dense and collisional H$_2$O and H$_2$ components \textit{cannot both be present} in Callisto's atmosphere, since as seen in Fig. \ref{fig:H_from_H2O_H2} redistribution of the H produced from H$_2$O via H$_2$+H collisions enhances this already large discrepancy. If $f$ is reduced to improve the agreement over the disk, the discrepancy between the H produced from H$_2$O and what is observed off the disk becomes even worse (Fig. \ref{fig:H_corona_final}b). Conversely, increasing $f$ to improve the agreement off the disk results in a discrepancy between the H produced from H$_2$O and what is observed over the disk (e.g., Figs. \ref{fig:H_corona_singlespecies}a and \ref{fig:H_from_H2O_H2}). This implies that a global source of H is needed so that such a sharp drop-off in local H production and the corresponding density off the disk does not occur, and the emissions on the disk are dominated by sunlight reflected from the surface rather than those from local atmospheric sources. Since, as described above, radiolytically produced H$_2$ is distributed globally at Callisto, we propose H$_2$ is this global source of H. Based on the observation over the disk the H$_2$O peak density must be $\lesssim$ 10$^8$ cm$^{-3}$ requiring that $f \lesssim 10^{-3}$ (e.g., Fig. \ref{fig:H_corona_final}). When the ice on Callisto's surface is assumed to be segregated into patches of bright ice and dark non-ice or ice-poor material and H$_2$O is assumed to primarily sublimate from the former at the local ice temperature, the spatial distribution is similar to that for the intimate mixture scenario, but the amount of H produced is insufficient to explain the observation (e.g., Fig.~\ref{fig:H_corona_singlespecies}d), even when the H abundance increases via collisions with other species. Therefore, H$_2$O is not the primary source of the H corona.

The constraint on the peak number density, $\lesssim$~10$^8$ H$_2$O cm$^{-3}$, corresponds to a peak sublimation flux of $\lesssim$~10$^{12}$ H$_2$O cm$^{-2}$ s$^{-1}$ (Fig. \ref{fig:SSL_vs_subflux}) and a peak radial column density of $\lesssim$~10$^{15}$~H$_2$O~cm$^{-2}$ (Fig. \ref{fig:H2O_RadColDens}). These constraints are significantly less than those implemented in the literature: applying a uniform surface temperature of 150 K, \cite{liang2005} derived a maximum number density of $\sim$2$\times$10$^9$ H$_2$O cm$^{-3}$ (Fig. 2 therein); applying a single day-side surface temperature profile with a $T_{max}$ = 165 K and a range of ice coverage of $\sim$60--73$\%$, \cite{vorburger2015} derived a maximum number density of $\sim$4$\times$10$^{10}$ H$_2$O~cm$^{-3}$ (Fig. 1 therein); applying a maximum temperature of 155 K \cite{hartkorn2017} derived a maximum column density of $\sim$3$\times$10$^{16}$ H$_2$O cm$^{-2}$ (Fig. 7 therein); and using the temperature distribution from \cite{hartkorn2017} and $f=10^{-1}$, CM21 derived a maximum sublimation flux of $\sim$10$^{13}$~H$_2$O~cm$^{-2}$~s$^{-1}$ (Fig. 1 therein), number density of $\sim$10$^9$ H$_2$O cm$^{-3}$ (Fig.~E.2 therein), and column density of $\sim$8$\times$10$^{15}$ H$_2$O cm$^{-2}$ (also Fig. E.2 therein). Despite the relatively low fluxes at Callisto considered here, sublimation still produces several orders of magnitude more H$_2$O than would sputtering of its surface, even when assuming the full magnetospheric particle flux reaches the surface (e.g., \citealt{vorburger2019}).

The most recent modeling efforts for Europa suggest sublimation fluxes of $\sim$10$^{11}$~H$_2$O~cm$^{-2}$~s$^{-1}$ (\citealt{plainaki2018} and references therein), whereas a large range of fluxes have been suggested for Ganymede’s ``dirty ice’’ (e.g., \citealt{leblanc2017}). While \cite{leblanc2017} suggested that sublimation from dirty ice could be lower than the lowest bound presented here ($\sim$10$^8$ H$_2$O cm$^{-2}$ s$^{-1}$, Fig. 2 therein), they also consider rates even higher than the upper bound considered here ($>$10$^{15}$~H$_2$O~cm$^{-2}$~s$^{-1}$, also Fig. 2 therein), which yields a column density similar to the largest estimate suggested here, $\sim$(3–-8)$\times$10$^{15}$ H$_2$O cm$^{-2}$. \cite{marconi2007} estimated an H$_2$O sublimation flux and corresponding column density at Ganymede within these bounds, $\sim$10$^{13}$~H$_2$O~cm$^{-2}$~ s$^{-1}$ and $\sim$6$\times$10$^{15}$ H$_2$O cm$^{-2}$, respectively. More recently, \cite{vorburger2021} attained a peak number density of $\sim$4$\times$10$^9$ cm$^{-3}$. Thus the constraints we estimated for H$_2$O are similar to what has been suggested at Europa and are within the broad range suggested at Ganymede, contrary to previous models that assumed more H$_2$O in Callisto's atmosphere.

\subsection{Radiolytically Produced H$_2$} \label{H2_Discussion}

Since, as described above, the morphology of H is similar to that of its parent species, the HST/STIS observation of a H corona at Callisto indicates a global source is required. As suggested in CM21, we confirm that a globally distributed H$_2$ component, in an atmosphere also containing H$_2$O and O$_2$, can qualitatively and quantitatively account for the observed H corona. Ignoring any ionospheric source, which is discussed below, the estimated average surface densities required to reproduce the observation are $n_{0, \mathrm{H_2}} \sim 10^8$ cm$^{-3}$ when hot H is solely produced via interactions with photons or $n_{0, \mathrm{H_2}} \sim$ 4$\times$10$^7$ cm$^{-3}$ when hot H is also produced via interactions with magnetospheric electrons using our simple estimate for the flux.

As in CM20 and CM21, we assume that the near-surface density of H$_2$, primarily produced via radiolysis, is essentially global as it is efficiently redistributed and permeates the porous regolith from which it thermally re-enters the atmosphere. Therefore, ignoring production in the atmosphere but accounting for dissociation and ionization, the simulated steady-state escape rate is indicative of the average production rate of H$_2$. In an H$_2$+H atmosphere, this varies from $\sim$(1.3--2.3)$\times$10$^{28}$ s$^{-1}$ (corresponding to $\sim$47--84 kg/s). When the O$_2$ component assumed here ($n_{0, \mathrm{O_2}} \sim$ 10$^9$ cm$^{-3}$; e.g., Fig.~\ref{fig:col_dens_final}) and the H$_2$O component based on our constraints are included, they affect the H$_2$ component and the H it produces, thereby reducing this range by about half to $\sim$(0.7--1.4)$\times$10$^{28}$ s$^{-1}$ (corresponding to $\sim$26--51~kg/s). The density of O$_2$ has been suggested to be larger than that considered here (e.g., \citealt{kliore2002}), and collisions with such a dense O$_2$ component would further reduce the H$_2$ escape rate. In addition, although absorption of reflected Lyman-$\alpha$ by O$_2$ is likely negligible (e.g., \citealt{roth2017b}, Fig.~2 therein), incoming photons could be absorbed by such a dense O$_2$ as we showed for H$_2$ (Fig.~\ref{fig:H2_Hproduction_reduction}), further affecting H production.

Consistent with CM21, H$_2$ escape rates do not linearly scale with $n_{0, \mathrm{H_2}}$ due to the influence of H$_2$+H$_2$ collisions. The production and corresponding density of H exhibit a similar trend due to the influence of H$_2$+H collisions: the more collisional the H$_2$ component, the more the H density profile is enhanced (e.g., Fig. \ref{fig:H_by_H2_LOSColDens}). This issue becomes even more complicated when other collisional components are included (e.g., a dense, near-surface O$_2$ component) because, as discussed earlier, they can affect the spatial distribution of H$_2$, which in turn affects the H produced.

While we determined that a direct surface source of H can not account for the observed corona (Section \ref{results:H_corona}), there are \textit{indirect} sources that we did not consider, which are outside the scope of this study. For example, below the H$_2$ exobase ionization can produce H$_2^+$, which in turn can collide with a neutral H$_2$ producing H$_3^+$ and a neutral H (H$_2^+$ + H$_2$ $\rightarrow$ H$_3^+$ + H), and on recombination H$_3^+$ can subsequently produce even more H (H$_3^+$ + $e$ $\rightarrow$ H + H + H) as well as recycle neutral H$_2$ (H$_3^+$ + $e$ $\rightarrow$ H$_2$ + H). These additional sources of energized H could reduce the required $n_{0, \mathrm{H_2}}$ to produce the observed H corona and the concomitant H$_2$ escape rates. However, as the H$_2$ atmosphere thins out due to the reduced surface density, then the H$_2$+H collisions will be less efficient at enhancing the H density via collisions. In addition, as the surface density decreases, so too do the H$_2$ exobase altitudes until the atmosphere is dominated by O$_2$ (e.g., CM21). When this is the case the ionosphere would be dominated by interactions with O$_2$, thereby limiting the contribution to the H corona and the H$_2$ escape rate. Thus, the decrease in H$_2$ density cannot be too much so that the H$_2$ exobase is still well above the other collisional components in Callisto's atmosphere. A self-consistent hybrid model combining the atmosphere, the ionosphere, and the surrounding plasma is required to investigate these interactions.

\subsection{\textit{Galileo} Plasma-Wave Observations} \label{discussion_gurnett}

\begin{figure}[t]
    \centering
    \subfloat[]{\includegraphics[width=0.42\textwidth]{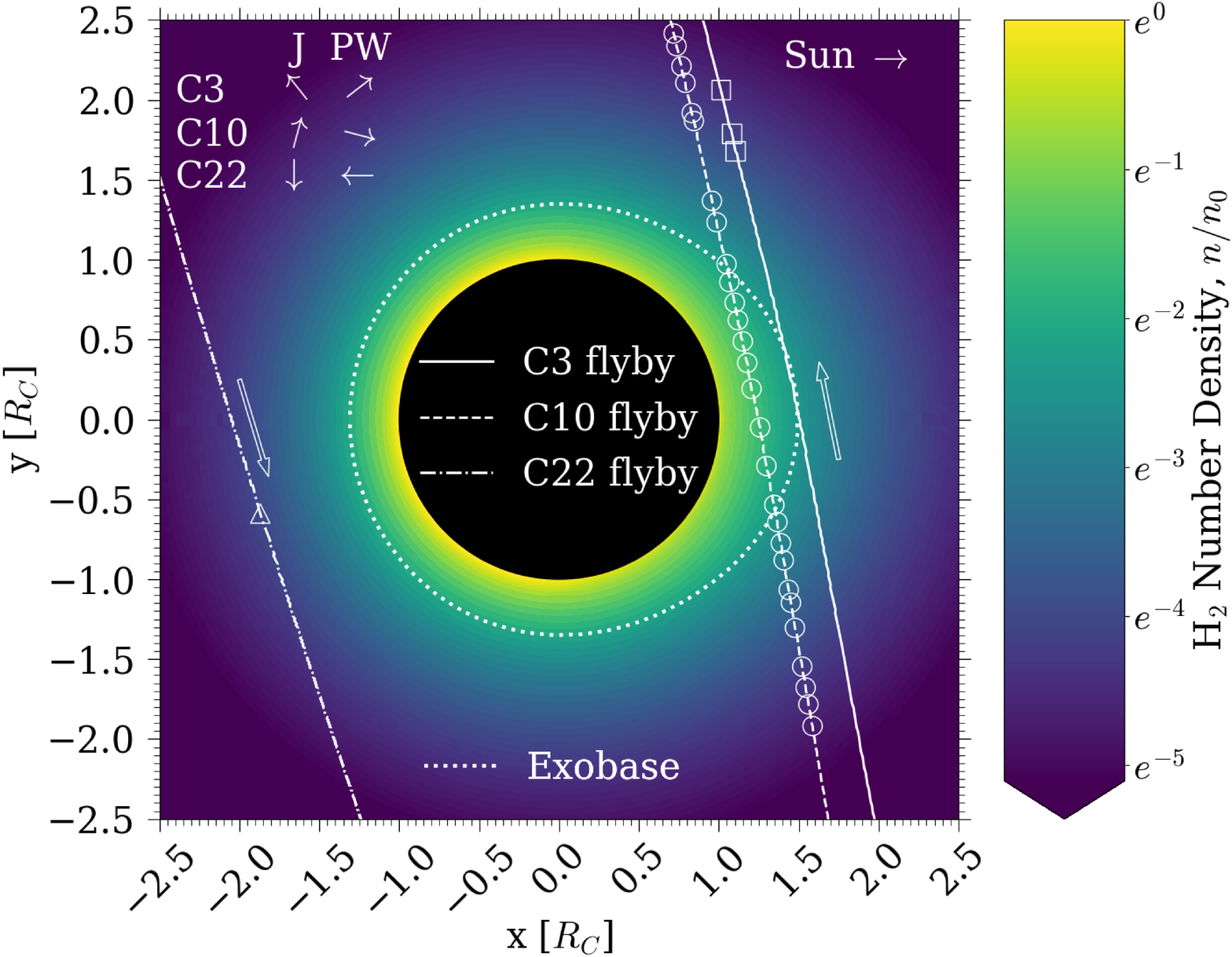}}
    \hfill
    \subfloat[]{\includegraphics[width=0.58\textwidth]{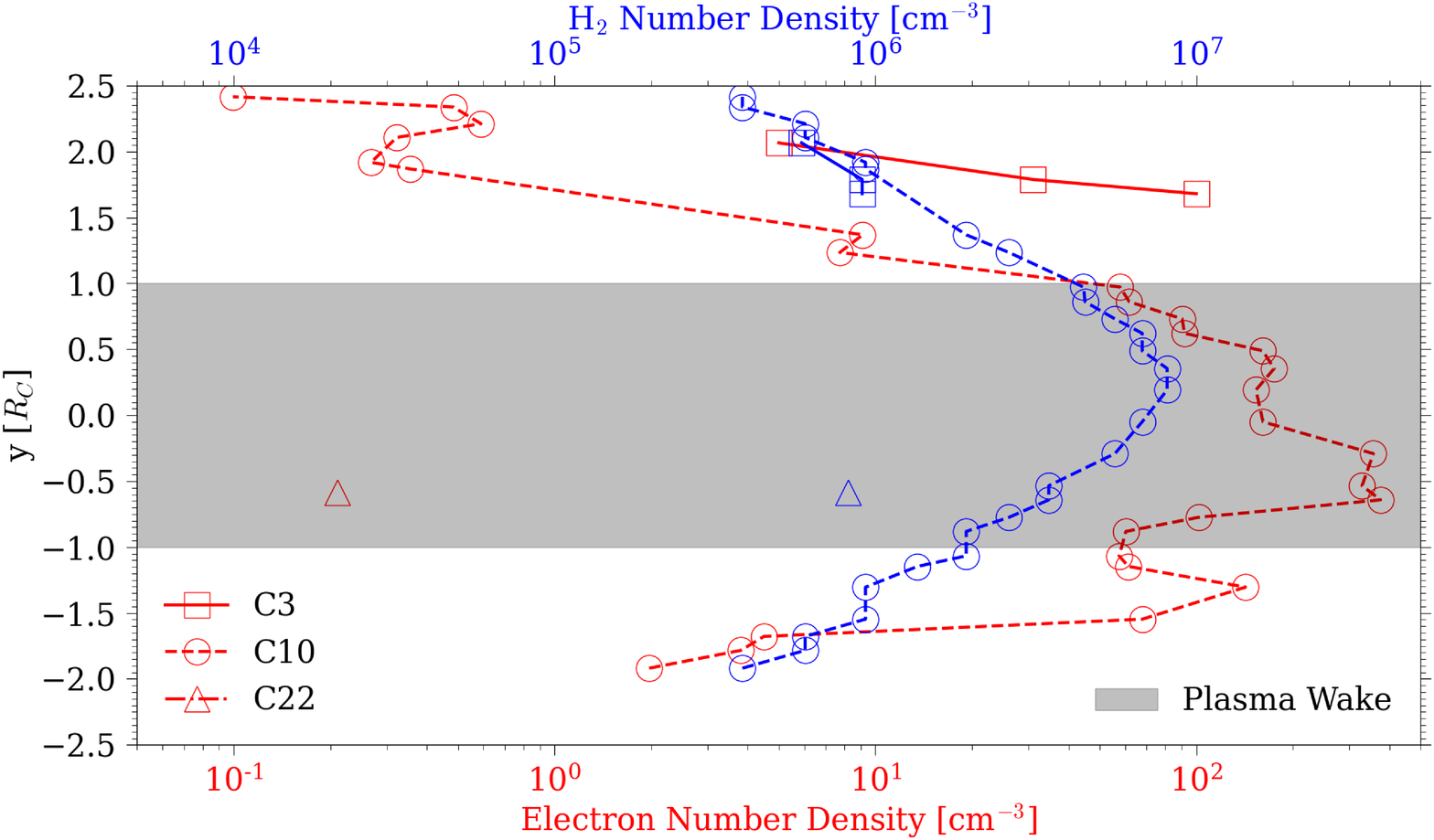}}
    \caption{(a) The C3 (solid line), C10 (dashed line), and C22 (dash-dotted line) flyby trajectories are superimposed onto the H$_2$ density profile where $n_{0,\mathrm{H_2}} \sim$ 4$\times$10$^7$ cm$^{-3}$, interactions with photons and magnetospheric electrons are considered as given in Table \ref{tab:reactions}, and an O$_2$ component with $n_{0, \mathrm{O_2}} \sim$ 10$^9$ cm$^{-3}$ is also present. In this Callisto-centered x-y coordinate system, the Sun is located to the right (positive x-direction); the points at which electron densities were observed are denoted by white squares, circles, and a triangle for the C3, C10, and C22 flybys, respectively; the exobase is represented by the dotted white line; and arrows illustrating the direction \textit{Galileo} flew by Callisto on the left (C22) and right (C3, C10) of the plot, as well as the general direction of Jupiter (``J'') and the plasma wake (``PW'') during these flybys in the upper left of the plot. (b) The electron densities (red lines; x-axis, \textit{bottom}) observed during the C3, C10, and C22 flybys compared to the corresponding simulated H$_2$ densities (blue lines; x-axis, \textit{top}) along the y-axis from (a). The shaded region roughly approximates the plasma wake region during all 3 flybys.} 
    \label{fig:Gurnett_flybys}
\end{figure}

\cite{gurnett1997} suggested that an extended atmospheric component was likely responsible for the enhanced electron densities observed by \textit{Galileo}’s plasma-wave instrument as the spacecraft crossed into the plasma sheet during the C3 flyby (e.g., Fig. \ref{fig:Gurnett_flybys}a). Since the observed electron densities at the relevant altitudes were orders of magnitude larger than the background magnetospheric electron density (c.f., Fig. \ref{fig:Gurnett_flybys}b and Table \ref{tab:background_plasma} in Appendix~\ref{app:phys_params}), they suggested it was produced by the ionization of a neutral gas implying ``a significant neutral atmosphere must exist around Callisto.’’ They also suggested it might be related to the atomic hydrogen cloud detected at Ganymede \citep{barth1997, feldman2000}. Since the C/A of this flyby, at 1129 km (1.47 $R_C$), is well above the altitudes of the presumably O$_2$-sourced ionospheric component deduced from the radio-wave deflections ($\lesssim$~50 km; \citealt{kliore2002}), an extended source is required, which we suggest is H$_2$. The subsequent flybys provided little clarity on the composition and spatial distribution of the suggested atmosphere \citep{gurnett2000}. Whereas the C22 flyby, with a C/A altitude of 2299 km (1.95 $R_C$), resulted in a maximum electron density of only $\sim$0.21 cm$^{-3}$ near C/A in the wake region, the C10 flyby with a C/A altitude of 535 km (1.22 $R_C$) confirmed the presence of a highly extended, enhanced component with peak densities orders of magnitude larger than that of the background Jovian magnetosphere plasma density at Callisto's orbit (c.f., Fig.~\ref{fig:Gurnett_flybys}b and Table \ref{tab:background_plasma} in Appendix \ref{app:phys_params}). Since relatively high plasma densities were measured within altitudes of 535 km (C10 flyby) and 1129 km (C3 flyby) but not at an altitude of 2299 km (C22 flyby), \cite{gurnett2000} suggested that the observations were indicative of a high altitude ionospheric distribution. In addition, as the C3 and C10 flybys occurred on the day-side but the C22 flyby occurred on the night-side (Fig. \ref{fig:Gurnett_flybys}a) they suggested its production was due to solar illumination.

Using a magneto-hydrodynamics model to describe the energetic plasma flow during the C10 flyby, \cite{liuzzo2016} considered only O$_2$ and CO$_2$ components and, as a result, required even larger neutral densities than those inferred by \cite{carlson1999} and \cite{kliore2002} to account for the plasma-wave observations. Because the densities of these components, as well as that of H$_2$O, are negligible at the relevant altitudes of the aforementioned flybys, and the local H density is much less than that of the H$_2$, we propose that the inferred electron densities, $n_e$, were determined by the local density of H$_2$, its ionization rate, and the electron loss rate. In Fig. \ref{fig:Gurnett_flybys}a we superimposed the C3, C10, and C22 trajectories onto the H$_2$ density profile for the case in which $n_{0, \mathrm{H_2}} \sim$ 4$\times$10$^{-7}$ cm$^{-3}$ in a multi-component atmosphere and interactions with both photons and magnetospheric electrons lead to ionization. The observation points, represented by circles, squares, and a triangle along the spacecraft trajectories are all very close to Callisto's orbital plane ($<$5$^\circ$~latitude) throughout the flybys with C/A ranging from 594--2299 km (1.22--1.95~$R_C$), and the direction of the spacecraft motion as well the plasma wake are indicated by arrows. The 3 trajectories are seen to sample a drop off of only a few H$_2$ scale heights, and several of the C10 measurements are taken below the H$_2$ exobase, suggesting the extended ionospheric processes described in Section \ref{H2_Discussion} could occur. Here we define the exobase at the altitudes where the Knudsen number, Kn~=~$\ell / H \sim 1$, with $\ell$ and $H$ the local mean free path and scale height of H$_2$, respectively.

Assuming the electrons produced from H$_2$ are primarily picked up and swept out of the region by the rapidly rotating Jovian field in a time $\tau_e$, then $n_e$ is roughly proportional to~$n_\mathrm{H2}$:

\begin{equation}
    n_\mathrm{H_2} \Sigma_i (\nu_{\mathrm{p}e} + \nu_{\mathrm{h}\nu})  \sim n_e / \tau_e.
    \label{eq:electron_lifetime}
\end{equation}

\noindent The data points for $n_e$ and $n_\mathrm{H_2}$ are represented in red and blue in Fig. \ref{fig:Gurnett_flybys}b, respectively; and $i$ indicates the sum over the ionizing processes induced via interactions with plasma electrons (p$e$) and photons (h$\nu$) in Table \ref{tab:reactions} in Appendix \ref{app:reactions} with a net ionization rate $\nu_{\mathrm{p}e} + \nu_{\mathrm{h}\nu} \sim$~7.4$\times$10$^{-8}$ s$^{-1}$. As will be discussed below, $\nu_{\mathrm{p}e}$ and $\tau_e$ can be highly variable, which would have a relatively small effect on the H$_2$ density profile, but a significant effect on the electron production rate. Assuming the net ionization rate is fixed, then from Fig.~\ref{fig:Gurnett_flybys}b $\tau_e$ would be roughly proportional to $n_e / n_\mathrm{H_2}$, and varies from $\sim$10$^{0}$--10$^{3}$ s. The shorter times are consistent with loss by pick-up and sweeping, which for a distance of $R_C$ is $\sim$10$^1$ s. The longer times in the more depleted regions are comparable to average half bounce times (the time an electron travels from Callisto to its mirror point and back), $\gtrsim$10$^2$ s for 100~eV electrons (e.g., \citealt{liuzzo2019}, Fig. 2a therein), suggesting the assumption of a uniform irradiation by the plasma electrons is a rough approximation.

In Fig. \ref{fig:Gurnett_flybys}b it is seen for C10 that as Galileo crosses the plasma wake region, the electron density very roughly scales with the H$_2$ density allowing for the expected plasma turbulence in the wake. Because our 2D model is axisymmetric about the line passing through the subsolar point, but the plasma flow is not, the differences in the profiles on either side of the wake are not surprising. As the ion gyroradii are on the order of or larger than Callisto's radius (\citealt{cooper2001}, Fig. 4 therein) and the magnetic field points downward, enhanced plasma densities prior to entering the wake and depleted densities on exiting are expected.

The observed electron densities for C10 differs from that for C22 in a similar region by about 3 orders of magnitude. Whereas the latter occurred near western elongation 4.31 $R_J$ below the plasma sheet and in Callisto's shadow, the former occurred near eastern elongation 2.45 $R_J$ below the plasma sheet and exposed to the Sun \citep{seufert2012}, which could contribute to the differences in observed electron densities. Contrary to this comparison, although the 3 measurements taken during the C3 flyby also very roughly scale with the H$_2$ density like those during the C10 flyby in a similar region, the electron densities for the former are about 1--2 orders of magnitude larger than those for the latter. Interestingly, the C3 flyby occurs when Callisto was farther from the plasma sheet, 3.24 $R_J$ and above it \citep{seufert2012}. Since both observations occur on the day-side with Callisto near eastern elongation and the average solar irradiance only varied by $\lesssim$10$\%$ (e.g., \citealt{schmutz2021}, Fig.~2 therein) in the 3 year interval between them, this comparison suggests that the plasma induced ionization dominates the difference in electron production. Indeed, as shown in Table \ref{tab:background_plasma} in Appendix \ref{app:phys_params}, the background plasma density was larger during the C3 flyby than that during the C10 flyby, with a temperature more efficient for ionization (e.g., \citealt{straub1996}). 

Since the C22 flyby occurred over the night-side of Callisto and the C3 and C10 flybys occurred over the day-side, photoionization could be a required source for the extended ionosphere and/or it could be a transient phenomena, varying temporally as well as spatially. Since the Jovian magnetodisk wobbles relative to Callisto's orbital plane by $\sim$10$^\circ$ and, as a result, Callisto moves in and out of the Jovian plasma sheet, it experiences magnetic distances between $\sim$26~$R_J$ ($d_{JC}$) and as large as $\sim$70~$R_J$ ($\sim$2.7 $d_{JC}$) (\citealt{paranicas2018}, Fig. 2 therein), resulting in a highly variable plasma environment. Thus a detailed model is needed to determine the highly variable source and loss rates in an extended H$_2$ atmosphere as Callisto moves in and out of the plasma sheet. Nevertheless, based on our results for a global H$_2$ atmosphere and the similar profiles illustrated in Fig. \ref{fig:Gurnett_flybys}b, we suggest that these observations are consistent with an extended H$_2$ atmosphere.

\subsection{Neutral H$_2$ Torus}

Since the H$_2$ that escapes from Callisto does not escape from the Jovian system and has a lifetime longer than Callisto’s orbital period (Table \ref{tab:torus}), a yet to be detected neutral H$_2$ torus will form (e.g., CM21). Using our simulation results for steady-state H$_2$ escape rates ($\Phi$) and speeds ($v_\mathrm{esc}$) as well as the corresponding lifetimes ($\tau$), we very roughly estimate the average torus density ($n_\mathrm{torus}$) co-rotating with Callisto using the analytical estimate from \cite{johnson1990} in Table \ref{tab:torus}. We also include the H$_2$ densities at Callisto's Hill sphere, $r_\mathrm{HS} \sim 0.7 R_J$, which is a rough upper bound for the peak densities in the torus.

\begin{table} [ht!]
    \centering
    \caption{Callisto's neutral H$_2$ torus.}
    \begin{tabular}{|c|c|c|}
        \hline
        & Upper Estimate ($a$) \tnote{a} & Lower Estimate ($b$) \tnote{b} \\
        \hline
        H$_2$ Surface Density,    & \multirow{2}{*}{1.0} & \multirow{2}{*}{0.4} \\
        {$n_0$ [($\times$10$^8$) cm$^{-3}$]}                                                 & & \\
        \hline
        Escape Rate,        & \multirow{2}{*}{1.4} & \multirow{2}{*}{0.68} \\
        {$\Phi_\mathrm{esc}$ [($\times$10$^{28}$) s$^{-1}$]}                & & \\
        \hline
        Escape Speed,       & \multirow{2}{*}{0.90} & \multirow{2}{*}{1.1} \\
        {$v_\mathrm{esc}$ ($c$) \tnote{c} [km s$^{-1}$]}                                    & & \\
        \hline
        Lifetime,           & \multirow{2}{*}{78 ($d$) \tnote{d}} & \multirow{2}{*}{8.4 ($e$) \tnote{e}} \\
        {$\tau$ [$t_\mathrm{orb}$]}                                         & & \\
        \hline
        Radial Extent, &  \multirow{2}{*}{12} & \multirow{2}{*}{14} \\
        {$\Delta R_\mathrm{torus}$ ($f$) \tnote{f} [$R_J$]}                 & & \\
        \hline           
        Scale Height,  &  \multirow{2}{*}{1.4} & \multirow{2}{*}{1.8} \\
        {$H_\mathrm{torus}$ ($g$) \tnote{g} [$R_J$]}                        & & \\
        \hline           
        Average Torus Density,  & \multirow{2}{*}{3.9} & \multirow{2}{*}{0.14} \\
        {$n_\mathrm{torus}$ ($h$) \tnote{h} [($\times$10$^2$) cm$^{-3}$]}    & & \\
        \hline
        Average Density at Hill Sphere,  & \multirow{2}{*}{9.8} & \multirow{2}{*}{4.4} \\
        {$n_\mathrm{H_2} (r_\mathrm{HS})$ [($\times$10$^2$) cm$^{-3}$]}    & & \\
        \hline           
    \end{tabular}
    \label{tab:torus}
    \begin{tablenotes}\footnotesize
    \item[a] ($a$) Simulation results where only interactions with photons are considered and an O$_2$ component with $n_{0, \mathrm{O_2}} \sim$ 10$^9$ cm$^{-3}$ is present.
    \item[b] ($b$) Simulation results where interactions with photons and an upper limit for magnetospheric electrons are considered and an O$_2$ component with $n_{0, \mathrm{O_2}} \sim$ 10$^9$ cm$^{-3}$ is present.
    \item[c] ($c$) For reference, the escape speed from Jupiter at $d_{JC}$ is $\sqrt{2 G M_J / (R_J + d_{JC})} \sim$ 11.4 km/s and the speed required to reach Callisto's Hill sphere from the surface is $\sqrt{2 G M_C (1/R_C - 1/r_\mathrm{HS})} \sim$ 2.39~km/s. Here $G =$ 6.674$\times$10$^{-11}$ m$^3$ kg$^{-1}$ s$^{-2}$ is the gravitational constant and $M_C$ = 1.076$\times$10$^{23}$ kg and $M_J$~=~1.898$\times$10$^{27}$ kg are the masses of Callisto and Jupiter, respectively.
    \item[d] ($d$) $\tau = 1 / (\Sigma_i \nu_{i, \mathrm{photons}})$, where $i$ represents the number of photochemical reactions considered and $\nu_\mathrm{photons}$ are the corresponding reaction rates from Table \ref{tab:reactions} in Appendix \ref{app:reactions} scaled to an ``average'' Sun (\citealt{huebner2015}; i.e., with solar activity = 0.5 from \url{https://phidrates.space.swri.edu/} and rates scaled to 5.2 AU).
    \item[e] ($e$) $\tau = 1 / ((\Sigma_i \nu_{i, \mathrm{photons}}) + (\Sigma_j \nu_{j, \mathrm{electrons}}))$, where $j$ represents the number of magnetospheric electron impact-induced reactions considered and $\nu_\mathrm{electrons}$ are the corresponding reaction rates from Table \ref{tab:reactions} in Appendix \ref{app:reactions}.
    \item[f] ($f$) $\Delta R_\mathrm{torus} = d_{JC} (4 v_\mathrm{esc}/v_C)$ \citep{johnson1990}.
    \item[g] ($g$) $H_\mathrm{torus} = d_{JC} (v_\mathrm{esc}/2 v_C)$ \citep{johnson1990}.
    \item[h] ($h$) $n_\mathrm{torus} = \Phi_\mathrm{esc} \tau$ / $V_\mathrm{torus}$ \citep{johnson1990}, where $V_\mathrm{torus} = 2 \pi d_{JC} (2 \Delta R_\mathrm{torus}) (2 H_\mathrm{torus})$ is the volume of the torus.
    \end{tablenotes}    
\end{table}

The lifetime, $\tau$, is estimated using average plasma parameters at Callisto's orbit, even though molecules with significant eccentricities experience a large range of plasma parameters. For example, the rough estimate for the radial extent, $\Delta R_\mathrm{torus} \sim 12–14~R_J$ (Table \ref{tab:torus}), implies that H$_2$ from Callisto could reach Ganymede’s orbit ($\sim$15 $R_J$), where the plasma densities are significantly larger, reducing $\tau$, as well as making it difficult to distinguish the sources. However, based on the rough estimate of the scale height, $H_\mathrm{torus} \sim 1.4–1.8~R_J$ (Table~\ref{tab:torus}), $n_\mathrm{torus}$ will have dropped off several orders of magnitude prior to reaching Ganymede's orbit. With much faster average escape speeds, $v_\mathrm{esc} \sim$ 12--16 km/s, and much smaller escape rates, $\Phi \sim$ (1.0--1.5)$\times$10$^{26}$ s$^{-1}$, any H component in the Callisto torus will be orders of magnitude smaller than that estimated for H$_2$.

Using our lowest estimate of $n_\mathrm{torus} \sim$14 cm$^{-3}$ (Table \ref{tab:torus}), Callisto's average H$_2$ torus density is on the order of Europa's neutral torus: \cite{lagg2003} and \cite{mauk2003} inferred average H$_2$ torus densities at Europa of $\sim$20--25 and $\sim$40 cm$^{-3}$, respectively; \cite{smyth2006} reported an average H$_2$ torus density at Europa of $\sim$90 cm$^{-3}$; and \cite{smith2019} reported an average neutral torus density of $\sim$27 cm$^{-3}$ for all species, but H$_2$ accounted for most of the particles in this estimate. \cite{kollmann2016} analyzed \textit{Galileo} Energetic Particles Detector (EPD) data at Europa's orbit over a range of magnetic latitudes to constrain the thickness of the local torus. Although results from their analysis were consistent with those of \cite{lagg2003}, since the source rate of energetic particles near Europa's orbit is not well known, they could only make a relatively broad constraint for equatorial H$_2$ of $\sim$1--410 cm$^{-3}$. From recent Juno observations, \cite{szalay2022} confirmed the presence of a Europa-genic neutral H$_2$ toroidal cloud, which is expected to be the primary cloud constituent \citep{smith2019}. However, since they were derived from H$_2^+$ pickup ion densities, they refrained from estimating the concomitant neutral H$_2$ densities in the toroidal cloud, and suggested more complex modeling is required to do so. The simulated H$_2$ neutral loss rates from Callisto presented here (Section \ref{H2_Discussion}) are larger than those derived for Europa by \cite{szalay2022}. Moreover, since the neutral lifetimes are much shorter at Europa, its toroidal cloud's spatial extent is more confined than at Callisto. Thus, a larger H$_2$ neutral toroidal cloud could form around Callisto, albeit a more detailed comparison is required.

\section{Conclusion} \label{conclusion}

We simulated Callisto’s thin atmosphere in 2D using the DSMC method \citep{bird1994} to determine the source of the H corona observed by HST/STIS at eastern elongation \citep{roth2017a}. Our results showed that H$_2$O is not the dominant source of H and that a roughly global extended source is required. Noting the presence of a near-surface atmosphere dominated by radiolytically produced O$_2$, we show that the source of the H corona is consistent with the concomitant H$_2$ atmospheric component.  This global H$_2$ atmosphere can also account for the electron densities observed by \textit{Galileo}'s plasma-wave instrument \citep{gurnett1997, gurnett2000}. Thus, we present the first evidence of H$_2$ in Callisto's atmosphere.

An upper limit to the H$_2$ surface source rate was determined when only interactions with photons are considered as sources of H, but a more realistic source rate was shown to depend strongly on the uncertain production of H by the magnetospheric electrons penetrating the extended H$_2$ atmosphere. It also depends on the not well constrained near-surface O$_2$ component and the ionosphere it can produce, the former can inhibit H$_2$ escape via collisions and the latter can affect the magnetospheric electron flux (e.g., \citealt{strobel2002}). Consistent with CM21, the reduction in steady-state escape rate does not scale linearly with the corresponding reduction in the H$_2$ surface density, which varied by a factor of 2.5 for the cases discussed, from $\sim$10$^8$ cm$^{-3}$ to $\sim$4$\times$10$^7$~cm$^{-3}$. These values were $\sim$10--25$\times$ less than that we assumed for O$_2$: $\sim$10$^9$ cm$^{-3}$, which produces a disk-averaged column density of $\sim$5$\times$10$^{15}$ cm$^{-2}$ (Fig. \ref{fig:col_dens_final}b), consistent with that suggested by \cite{cunningham2015} 3.4$^{+2.0}_{-1.8}\times$10$^{15}$ cm$^{-2}$.

Simulations of the production of H from sublimated H$_2$O for two significantly different models of the surface ice were considered in which the ice and dark non-ice or ice-poor material are intimately mixed or segregated into patches. Neither scenario reproduces the observed morphology of the H corona. Based on the often used ``Intimate Mixture’’ model the peak H$_2$O number density is constrained to $\lesssim$10$^8$ cm$^{-3}$, indicating that previous models of Callisto's atmosphere overestimated the amount of gas phase H$_2$O by 1--2 orders of magnitude. Our simulation results place upper limits on peak H$_2$O production and density in Callisto's atmosphere, but cannot distinguish which of the two sublimation scenarios we considered is more likely--something that forthcoming spacecraft observations can distinguish. However, based on the visual evidence (e.g., \citealt{spencer1984, spencer1987b, moore1999}), we suggest that H$_2$O sublimation at Callisto is more likely to occur in a manner similar to that described by the ``Segregated Patches’’ scenario. Other possible direct sources for the H were considered, such as proton charge-exchange and direct sputtering of H, but are insufficient to reproduce the observation.

Although no neutral torus has yet been detected at Callisto’s orbit, based on the range of H$_2$ escape rates, escape speeds, and estimated lifetimes, a range of approximate average densities of the neutral H$_2$ torus co-rotating with Callisto is given. Comparing source rates to those estimated at Ganymede \citep{marconi2007, leblanc2017} and Europa (\citealt{szalay2022} and references therein), in combination with longer local lifetimes, a potentially detectable H$_2$ torus could form around Callisto. Further, the constraints we estimated for H$_2$O sublimation fluxes are similar to what has been suggested at Europa and are within the broad range suggested at Ganymede, contrary to previous models that assumed much larger fluxes at Callisto. The results presented here also suggest that the role of H$_2$ versus H$_2$O as a source of an H corona should be re-examined at Europa (e.g., \citealt{roth2017b}) and Ganymede (e.g., \citealt{barth1997, feldman2000, alday2017}) since their morphologies were similar to that at Callisto but the primary source of the H was assumed to be sublimated H$_2$O. The possibility of H$_2$ being the primary source of their H coronae can affect our understanding of the weathering of their icy surfaces and production of their atmospheres.

Future work is needed to examine reactions in the regolith, the neutral atmosphere, and the ionosphere to determine how much H$_2$ comes from water ice and from hydrocarbons and hydrated minerals in the dark non-ice surface materials, as well as to better constrain the O$_2$ component in Callisto's atmosphere. DSMC simulations, similar to those presented here, can be used to determine how much O$_2$ is required to reproduce the observed O corona \citep{cunningham2015} as well as any contribution from H$_2$O (e.g., \citealt{roth2021b}), thereby better constraining their source rates and surface densities. More important, a self-consistent model coupling Callisto's atmosphere and ionosphere to the local plasma environment is needed to account for the possible ionospheric sources of H described in Section \ref{H2_Discussion}.

Herein we have shown that in a four component (H$_2$O, H$_2$, O$_2$, H) collisional atmosphere at Callisto subject to interactions with solar photons and magnetospheric electrons the morphology of the observed H is reproduced primarily via dissociation of H$_2$ with an average surface density of $n_{0, \mathrm{H_2}} \sim$ 4$\times$10$^7$ cm$^{-3}$, with a minor contribution of H from sublimated H$_2$O with a peak number density $\lesssim$10$^8$ cm$^{-3}$ (Figs. \ref{fig:H_corona_final}--\ref{fig:col_dens_final}). The principal findings of this study -- the first evidence of an appreciable H$_2$ component and constraints on the sublimation of H$_2$O in Callisto's enigmatic atmosphere -- are exciting as they have implications for other icy satellites, lead to suggestions for possible spacecraft confirmation, and open the door for several new studies going forward.

\newpage

\appendix
\counterwithin{figure}{section}
\section*{Appendix}

\setcounter{table}{0}
\renewcommand*\thetable{\Alph{section}.\arabic{table}}

\section{DSMC and LOS Grids} \label{app:grid}

\begin{figure}[h!]
    \centering
    \includegraphics[width=\textwidth]{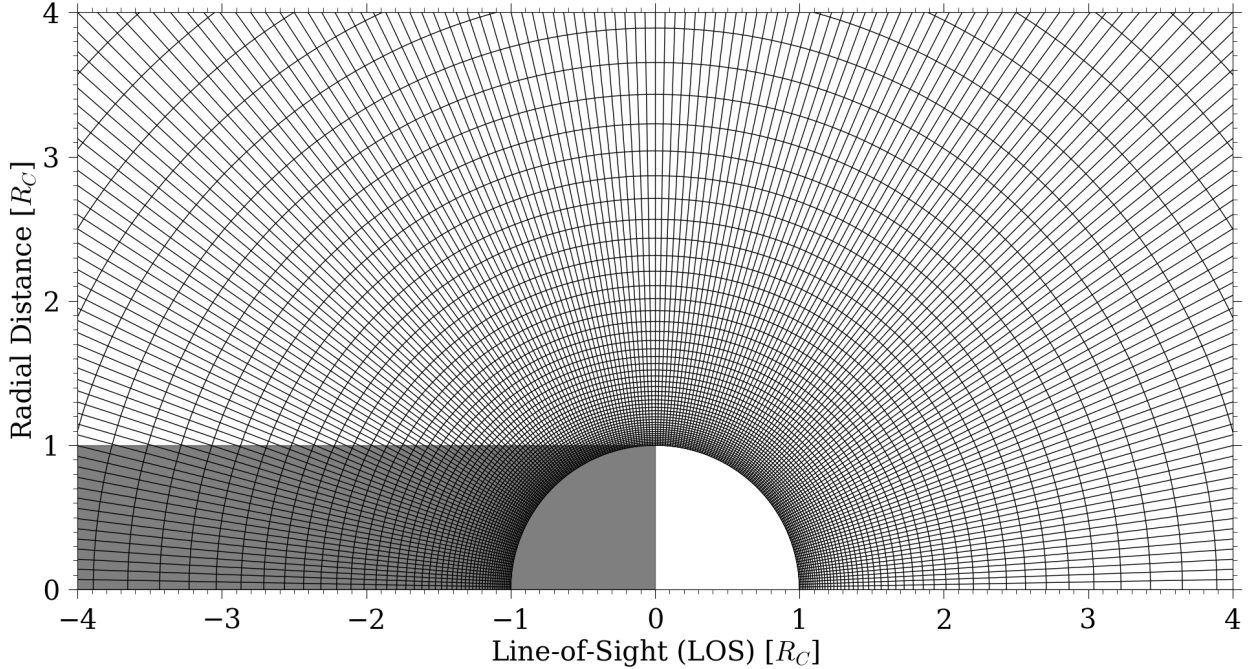}
    \caption{Example of a 2D DSMC grid used here. The lower and upper boundaries of the radial cells are set to Callisto's surface, $r_0 = R_C = 2410$ km, and the Hill sphere, $r_\mathrm{max} \sim 20.8~R_C$ ($r_\mathrm{max}-r_0 \sim$ 48,000 km), respectively. We refer the reader to CM21 (Appendix B therein) for how the extent of radial cells are determined The subsolar latitude (SSL) axis extends from the subsolar point (SSL = 0$^\circ$) to the anti-solar point (SSL = 180$^\circ$) with uniform angular increments $\Delta$SSL = 1$^\circ$, and the domain is symmetric about the axis passing through these points. The x-axis of this figure represents the LOS from the Sun to Callisto, and extends from $-r_\mathrm{max} \rightarrow +r_\mathrm{max}$, where positive and negative values represent the foreground and background of Callisto, respectively, with the center of Callisto at x~=~0. The y-axis of this figure is the radial distance perpendicular to the LOS, and extends from $0 \rightarrow r_\mathrm{max}$, with the center of Callisto at y = 0. The Sun is located to the right of the figure (positive x-axis). The shaded region on and off the disk represents the night-side and Callisto's shadow, respectively.}
    \label{fig:DSMC_grid}
\end{figure}

\newpage

\begin{figure}[h!]
    \centering
    \includegraphics[width=\textwidth]{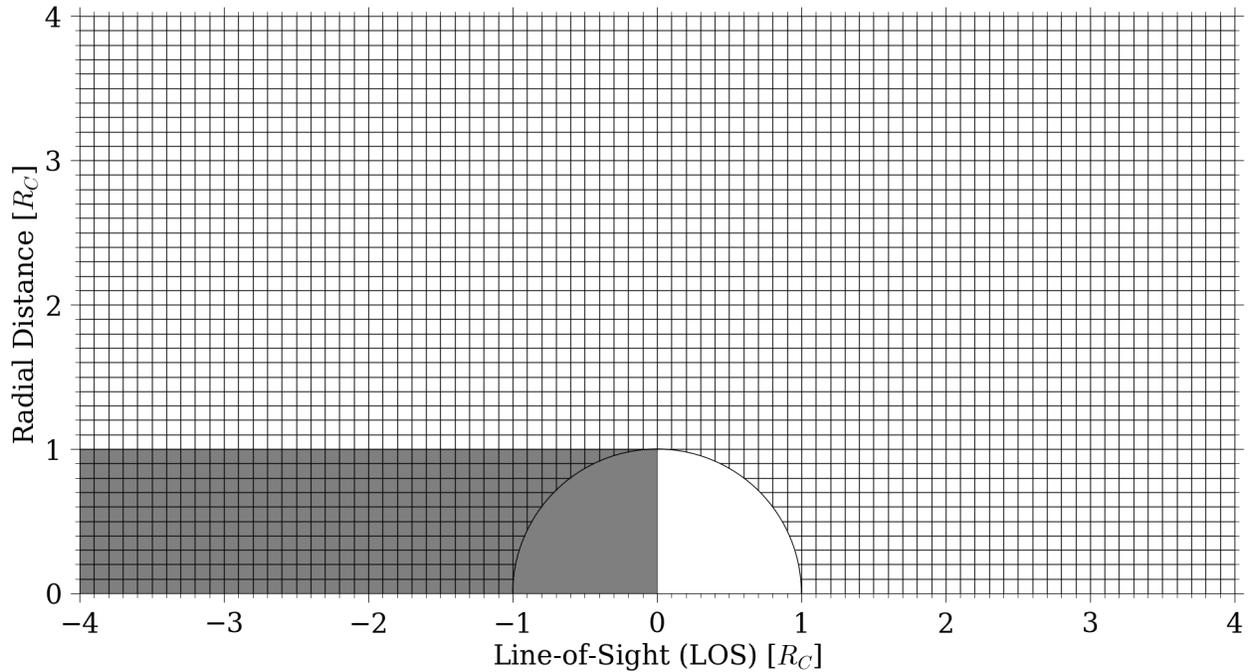}
    \caption{The 2D line-of-sight (LOS) grid used here to calculate H LOS column densities (Figs. \ref{fig:H_by_H2O_LOSColDens}--\ref{fig:H_by_H2_LOSColDens} and \ref{fig:H_corona_singlespecies}--\ref{fig:H_corona_final}) and photoabsorption due to opacity (Fig. \ref{fig:H2_Hproduction_reduction}). The x-axis of this figure represents the LOS from the Sun to Callisto, and extends from $-r_\mathrm{max} \rightarrow +r_\mathrm{max}$, where positive and negative values represent the foreground and background of Callisto, respectively, with the center of Callisto at x = 0. The y-axis of this figure is the radial distance perpendicular to the LOS, and extends from $0 \rightarrow r_\mathrm{max}$, with the center of Callisto at y~=~0. The increments for both axes are 0.1~$R_C$. A particle, $_p$, is located within a certain cell if $r_p \cos(\mathrm{SSL}_p)$ and $r_p \sin(\mathrm{SSL_p})$ are within the cell's bounds along the x- and y-axes, respectively. The Sun is located to the right of the figure (positive x-axis). The shaded region on and off the disk represents the night-side and Callisto's shadow, respectively.}
    \label{fig:LOS_grid}
\end{figure}

\newpage

\section{Photochemical and Electron Impact-Induced Reaction Rates} \label{app:reactions}

\begin{table}[h!]
    \centering
    \caption{Dissociation, ionization, and dissociative ionization rates via interactions with photons and magnetospheric electrons in Callisto's atmosphere.}
    \begin{tabular}{|ll|c|c|}
        \hline
        \multicolumn{2}{|c|}{Reaction} & Rate [s$^{-1}$] & Excess Energy [eV] \\
        \hline
        1)  & H$_2$O + h$\nu$ $\rightarrow$ H + OH              & 6.51$\times$10$^{-7}$     & 4.04  \\
        2)  & H$_2$O + h$\nu$ $\rightarrow$ H + H + O           & 7.06$\times$10$^{-8}$     & 0.697 \\
        3)  & H$_2$O + h$\nu$ $\rightarrow$ H$_2$ + O           & 5.47$\times$10$^{-8}$     & 3.94  \\
        4)  & H$_2$O + h$\nu$ $\rightarrow$ H$_2$O$^+$ + $e$    & 3.06$\times$10$^{-8}$     & 15.2  \\
        5)  & H$_2$O + h$\nu$ $\rightarrow$ H + OH$^+$ + $e$    & 5.58$\times$10$^{-9}$     & 23.2  \\
        6)  & H$_2$O + h$\nu$ $\rightarrow$ OH + H$^+$ + $e$    & 1.50$\times$10$^{-9}$     & 30.5  \\
        7)  & H$_2$O + h$\nu$ $\rightarrow$ H$_2$ + O$^+$ + $e$ & 8.17$\times$10$^{-10}$    & 39.8  \\
        8)  & H$_2$ + h$\nu$ $\rightarrow$  H$_2^+$ + $e$       & 4.25$\times$10$^{-9}$     & 7.17  \\
        9)  & H$_2$ + h$\nu$ $\rightarrow$ H + H                & 4.03$\times$10$^{-9}$     & 8.22  \\
        10) & H$_2$ + h$\nu$ $\rightarrow$ H + H$^*$            & 3.04$\times$10$^{-9}$     & 0.488 \\
        11) & H$_2$ + h$\nu$ $\rightarrow$ H + H$^+$ + $e$      & 1.03$\times$10$^{-9}$     & 27.0  \\
        \hline
        12) & H$_2$O + $e$ $\rightarrow$ H + OH + $e$           & 1.51$\times$10$^{-7}$     & 4.04  \\
        13) & H$_2$O + $e$ $\rightarrow$ H$_2$O$^+$ + 2$e$      & 9.65$\times$10$^{-8}$     & 15.2  \\
        14) & H$_2$O + $e$ $\rightarrow$ H + OH$^+$ + 2$e$      & 3.08$\times$10$^{-8}$     & 23.2  \\
        15) & H$_2$O + $e$ $\rightarrow$ OH + H$^+$ + 2$e$      & 2.65$\times$10$^{-8}$     & 30.5  \\
        16) & H$_2$O + $e$ $\rightarrow$ H + H + O + $e$        & 1.10$\times$10$^{-9}$     & 0.697 \\
        17) & H$_2$ + $e$ $\rightarrow$ H$_2^+$ + 2$e$          & 6.38$\times$10$^{-8}$     & 7.17  \\
        18) & H$_2$ + $e$ $\rightarrow$ H + H + $e$             & 1.29$\times$10$^{-8}$     & 0.488 \\
        19) & H$_2$ + $e$ $\rightarrow$ H + H$^+$ + 2$e$        & 5.88$\times$10$^{-9}$     & 27.0  \\
        \hline        
    \end{tabular}
    \label{tab:reactions}
    \begin{tablenotes}\footnotesize
    \item[a]Reactions 1--11: From \cite{huebner2015} for an ``active'' Sun (i.e., assuming solar maximum) and scaled to 5.2 AU, ignoring possible absorption with depth into the atmosphere. The Sun was near solar maximum at the time of the original observation of Callisto from which the H corona was detected \citep{strobel2002}, December 2001.
    \item[b]Reactions 12--19: Assume the electron number density, $n_e$, and temperature, $k_B T_e$, derived from Voyager measurements by \cite{neubauer1998}: $n_e = 1.1$ cm$^{-3}$ and $k_B T_e = 100$ eV.
    \item[c]Reactions 12--17, 19: Assume same excess energies as analogous photochemical reactions.
    \item[d]Reaction 18: Excess energy from Reaction 10 (e.g., \citealt{tseng2013}; see discussion in Section \ref{sect:reactions}).
    \item[e]Reaction 12: Rate based on cross section data from \cite{harb2001}.
    \item[f]Reactions 13--15: Rates based on cross section data from \cite{itikawa2005}.
    \item[g]Reaction 16: Rate based on cross section data from \cite{kedzierski1998}.
    \item[h]Reaction 18: Rate based on cross section data from \cite{scarlett2018}.
    \item[i]Reaction 17, 19: Rates based on cross section data from \cite{straub1996}.
    \end{tablenotes}
\end{table}

\newpage

\section{Temperature-Dependent Residence Time for H$_2$O} \label{app:temp_res_time}

\begin{figure}[ht]
    \centering
    \includegraphics[width=\textwidth]{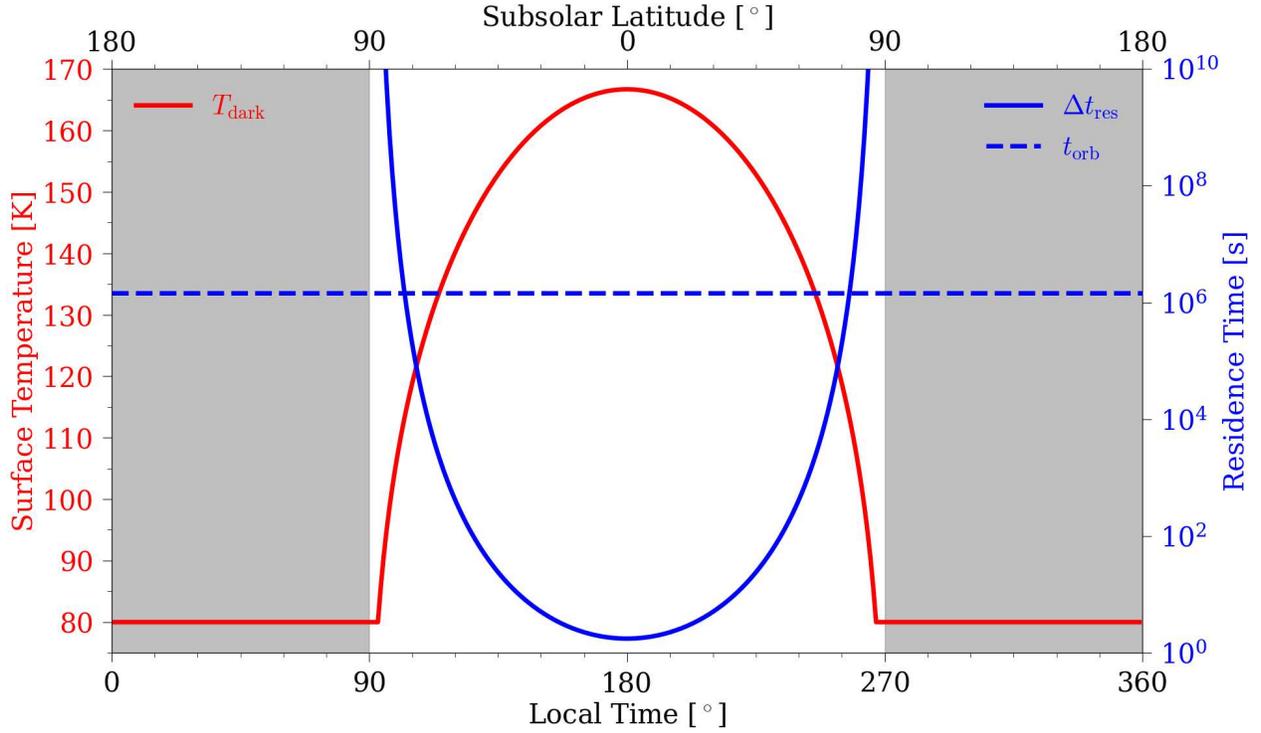}
    \caption{Temperature-dependent residence time for H$_2$O molecules on the non-ice or ice-poor material, $\Delta t_\mathrm{res}$ (solid blue line, right y-axis) as a function of the corresponding temperature distribution, $T_\mathrm{dark}(\mathrm{SSL})$ (solid red line, left y-axis), in the Segregated Patches scenario. For reference we also include Callisto's orbital period, $t_\mathrm{orb}$ = 1.44$\times$10$^6$ s (dashed blue line). The grey background represents the night-side.}
    \label{fig:T0_vs_TempResTime}
\end{figure}

The binding energy and lattice vibration frequency of H$_2$O used to derive $\Delta t_\mathrm{res}$ are taken from ice-related experiments \cite{sandford1988, sandford1990}, but are used here for H$_2$O molecules on Callisto's non-ice or ice-poor surface material. This same implementation of applying $\Delta t_\mathrm{res}$ for H$_2$O molecules landing on non-ice material has been done at the Moon for H$_2$O hopping away from cold traps and desorbing from Lunar regolith \citep{stewart2011}. Thus, although simulating the ensuing processes which occur after H$_2$O molecules return to the non-ice or ice-poor patches can be improved in the future, this approach is reasonable for a first order approximation as has been similarly applied in the literature. Finally, since these DSMC simulations can be run for several Callisto orbital periods, $t_\mathrm{orb}$~=~1.44$\times$10$^6$~s, to reduce statistical noise, if $\Delta t_\mathrm{res} > t_\mathrm{orb}$ then the particle is removed from the simulation. Removing these particles has a negligible affect on the overall distribution since relatively few particles migrate to such regions of the surface where $\Delta t_\mathrm{res} > t_\mathrm{orb}$. Moreover, consistent with CM21, steady-state for the SP - Ice Return simulations are reached in a fraction of $t_\mathrm{orb}$.

\newpage

\section{Callisto's Physical and Local Plasma Parameters} \label{app:phys_params}

\setcounter{table}{0}

\begin{table} [h!]
    \centering
    \caption{Various physical parameters of Callisto.}
    \begin{tabular}{|c|c|}
        \hline
        \multicolumn{2}{|c|}{Physical Parameters [units]} \\
        \hline
        {Radius, $R_C$ [km]} & {2,410} \\
        \hline
        {Mass, $M_C$ [kg]} & {1.08$\times$10$^{23}$} \\
        \hline 
        {Distance from Jupiter, $d_{JC}$ [$R_J$ $^a$ \tnote{a}]} & {26.3} \\
        \hline
        {(Average) Distance from Sun [AU]} & {5.20} \\
        \hline
        {Hill sphere, $r_\mathrm{HS}$ $^b$ \tnote{b} [$R_C$ ($R_J$)]} & {20.8 (0.701)} \\
        \hline
        {Orbital Period, $t_\mathrm{orb}$ [s (days)]} & {1.44$\times$10$^6$ (16.7)} \\
        \hline
        {Orbital Velocity, $v_C$ [km s$^{-1}$]} & {8.20} \\
        \hline
    \end{tabular}
    \label{tab:Callisto_params}
    \begin{tablenotes}\footnotesize
    \item[a] $^a$ $R_J = 71,492$ km is the radius of Jupiter.
    \item[b] $^b$ $r_\mathrm{HS} = \left( \frac{M_C}{3 M_J} \right)^{1/3} d_{JC}$, where $M_J$ = 1.898$\times$10$^{27}$ kg is the mass of Jupiter.
    \end{tablenotes}
\end{table}

\begin{table} [h!]
    \centering
    \caption{Background plasma density and ion temperature estimates taken from Tables 3.1 and 3.2 in \cite{seufert2012} for \textit{Galileo} C3, C10, and C22 flybys.}
    \begin{tabular}{|c|c|c|c|}
        \hline
        \multirow{2}{*}{Flyby} & Distance from & Background Plasma & Ion \\
        & Plasma Sheet$^a$ \tnote{a} [$R_J$] &  Density [cm$^{-3}$] & Temperature$^b$ \tnote{b} [eV] \\
        \hline
        C3 & 3.24 & 0.12 & 111 \\
        \hline
        C10 & -2.45 & 0.04 & 964 \\
        \hline
        C22 & -4.31 & 0.03 & 1,030 \\
        \hline
    \end{tabular}
    \label{tab:background_plasma}
    \begin{tablenotes}\footnotesize
    \item[a] $^a$ Positive and negative values indicate Callisto was above and below the plasma sheet, respectively.
    \item[b] $^b$ Ion temperature in the background plasma will be comparable to that of the electrons.
    \end{tablenotes}
\end{table}

\begin{table} [h!]
    \centering
    \caption{Proton parameters taken from \cite{vorburger2019} and references therein.}
    \begin{tabular}{|c|c|c|}
        \hline
        Parameter [units] & Thermal & Energetic \\
        \hline
        Number Density [cm$^{-3}$] & (10$^{-3}$--10$^{-1}$)$^a$ \tnote{a} & (10$^{-4}$--10$^{-2}$)$^a$ \\
        \hline
        Temperature [keV] & 0.2 & 20\\
        \hline
        Speed [km/s] & 192$^b$ \tnote{b} & 3,910$^c$ \tnote{c} \\
        \hline
    \end{tabular}
    \label{tab:V19_proton}
    \begin{tablenotes}\footnotesize
    \item[a] $^a$ Lower and upper bounds roughly represent minimum and maximum values of the local proton densities, respectively.
    \item[b] $^b$ Average relative speed between the local plasma azimuthal velocity and Callisto's orbital velocity \citep{kivelson2004}.
    \item[d] $^c$ $\sqrt{8 k T / \pi / m_\mathrm{H+}}$, where $k T$ = 20 keV.
    \end{tablenotes}
\end{table}

\newpage

\begin{table} [h!]
    \centering
    \caption{Temperature-dependent charge-exchange cross-sections taken from \cite{rinaldi2011} and references therein.}
    \begin{tabular}{|l|c|c|}
        \hline
        \multirow{2}{*}{Reaction} & Temperature & Cross-Section$^a$ \tnote{a} \\
        & [keV] &  ($\times$10$^{16}$) [cm$^2$] \\
        \hline
        H$^+$ + H $\rightarrow$ H + H$^+$           & 0.1--10 & 4--30       \\
        \hline
        H$^+$ + O $\rightarrow$ H + O$^+$           & 0.1--10 & 3.4--70     \\
        \hline
        H$^+$ + H$_2$ $\rightarrow$ H + H$_2^+$     & 0.1--10 & 0.3--10     \\
        \hline
        H$^+$ + O$_2$ $\rightarrow$ H + O$_2^+$     & 0.5--10 & 5.6--14     \\
        \hline
        H$^+$ + H$_2$O $\rightarrow$ H + H$_2$O$^+$ & 0.5--5 & 12.4--19.8   \\
        \hline
        H$^+$ + CO$_2$ $\rightarrow$ H + CO$_2^+$   & 0.1--10 & 7--16       \\
        \hline
    \end{tabular}
    \label{tab:CEX}
    \begin{tablenotes}\footnotesize
    \item[a] $^a$ Upper and lower bounds represent minimum and maximum values estimated over the temperature range given, which are not necessarily related to the upper and lower bounds of the temperatures.
    \end{tablenotes}
\end{table}

\section{Wavelength-Dependent Photoabsorption Cross Sections and Photochemical Reaction Rates} \label{app:wavelengths}

\begin{figure}[h!]
    \centering
    \includegraphics[width=\textwidth]{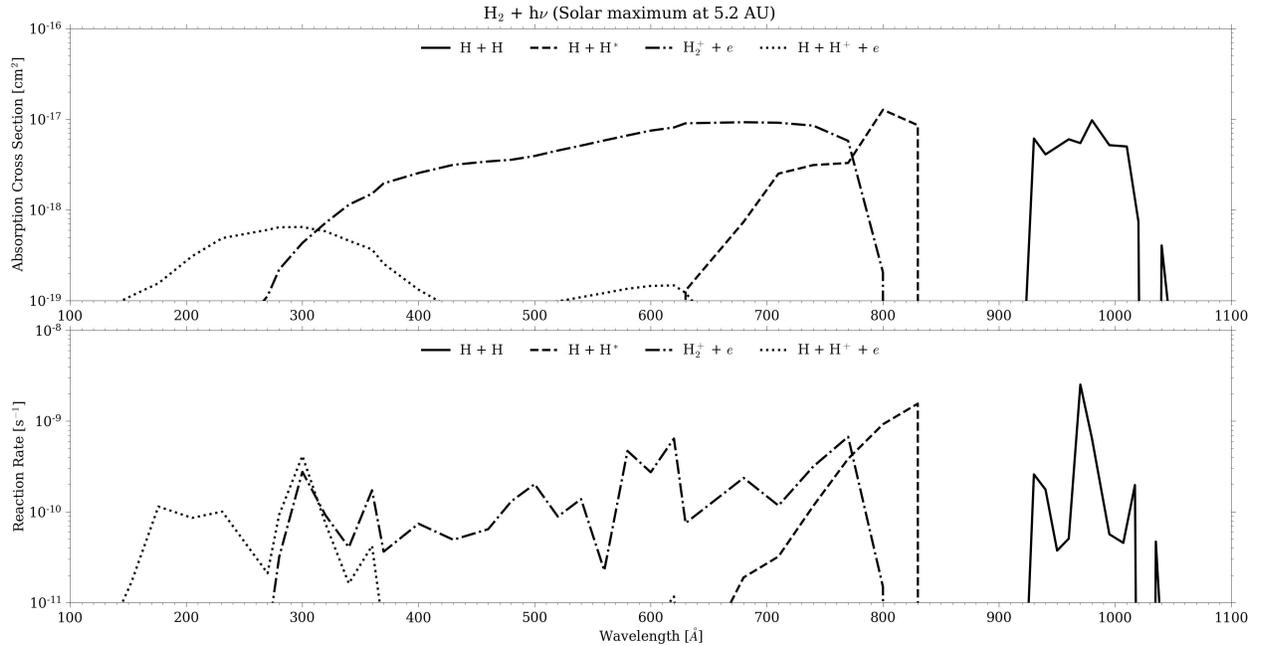}
    \caption{Photoabsorption cross sections (\textit{top}) and photochemical reaction rates (\textit{bottom}) as a function of  wavelength for H$_2$ assuming solar maximum (\citealt{huebner2015}; \url{https://phidrates.space.swri.edu/}) scaled to 5.2 AU.}
    \label{fig:H2_absorbxsec_reactionrates}
\end{figure}

\newpage

\section{Statistical Noise in DSMC Results}\label{app:stat_noise}

\begin{figure}[h!]
    \centering
    \includegraphics[width=\textwidth]{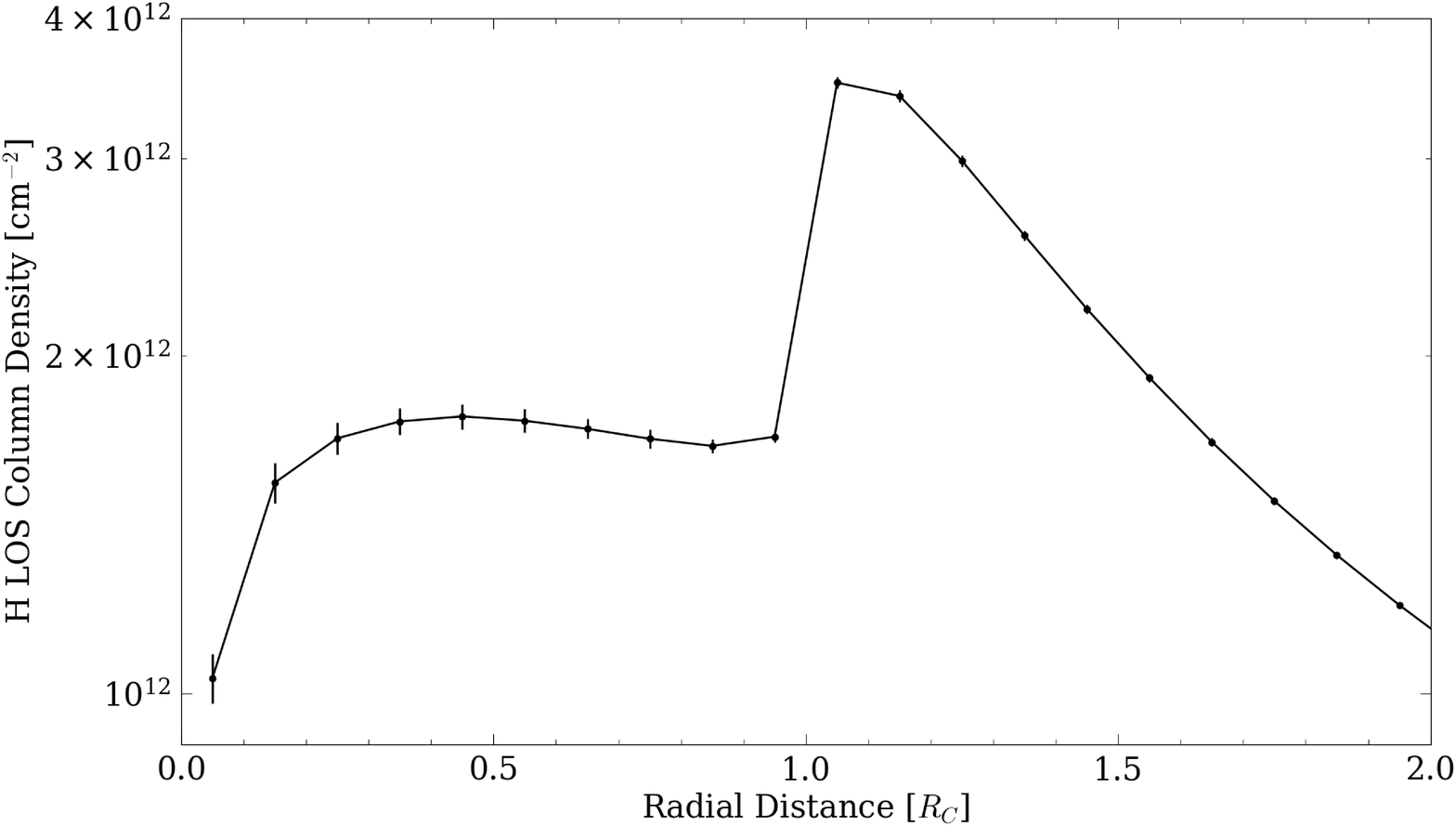}
    \caption{Steady-state result for H line-of-sight (LOS) column density in an H$_2$+H atmosphere illustrated by a solid black line with the standard deviation between averaged values at each data point shown by small vertical black lines. The dip in LOS column density at the center of the disk is an artifact of using radial bins for the density simulations and azimuthal bins for the LOS estimate. After reaching steady-state, this simulation was run for an additional 10$^6$ time-steps, with samples taken every $10^3$ time-steps and averages of those samples taken every 10$^5$ time-steps. As can be seen, the largest uncertainty is over the disk, where the LOS bins are the smallest, and with increasing distance from the disk and increasing LOS bin size, the uncertainty diminishes. These small, negligible fluctuations in LOS column density over the disk do not affect the comparison of the corresponding coronal brightness to the HST data because the atmospheric signals overlap with the sunlight reflected from the surface, which dominates the local emissions (Section \ref{results:H_corona}). As discussed in Section \ref{forward_model}, off the disk, where these fluctuations are seen to be even smaller, the sunlight reflectance is negligible and the comparison to the HST data are the most reliable for constraining the H corona.}
    \label{fig:LOS_Column_StdDev}
\end{figure}

\newpage

\section*{Acknowledgments}

Support for this research was provided by NASA GSFC/EIMM within NASA’s Planetary Science Division Research Program. The authors acknowledge the two anonymous reviewers whose comments and suggestions greatly improved the manuscript, as well as F. Leblanc, J. Chaufray, and L. Leclercq for their guidance in implementing the collision scheme from \cite{lewkow2014}. The figures for the \textit{Voyager} and \textit{Galileo} flyby trajectories were made using SPICE kernels \citep{acton1996, acton2018}. Finally, the simulations presented herein were carried out on the High Performance Computing resources at New York University Abu Dhabi.

\section*{Open Research}

The data used from the various models presented here to produce Figs. \ref{fig:H2O_RadColDens}--\ref{fig:Gurnett_flybys} can be accessed from \cite{carberrymogan2022}.

\bibliography{References.bib}

\end{document}